\documentclass[a4paper,11pt]{article}
\usepackage{jcappub}
\usepackage{lineno}

\def \dd {\mathrm{d}} 

\newcommand{\beq}{\begin{equation}}
\newcommand{\eeq}{\end{equation}}
\newcommand{\bea}{\begin{eqnarray}}
\newcommand{\eea}{\end{eqnarray}}


\newcommand\ees{\end{eqnarray}}
\newcommand\bees{\begin{eqnarray}}

\newcommand{\aM}{\alpha_{\text{M}}}
\newcommand{\aMzero}{\alpha_{\text{M0}}}
\newcommand{\aT}{\alpha_{\text{T}}}
\newcommand{\aTzero}{\alpha_{\text{T0}}}
\newcommand{\DTe}{\Delta t_{\text{e}}}
\newcommand{\DTa}{\Delta t_{\text{a}}}

\usepackage[english]{babel}

\usepackage[utf8]{inputenc}
\usepackage[T1]{fontenc}
\usepackage{ucs}
\usepackage{microtype}
\usepackage{color}
\usepackage{lettrine}
\usepackage{soul}
\usepackage{url}
\usepackage{hyperref}

\usepackage{xspace}
\usepackage{lastpage}
\usepackage{stmaryrd}

\usepackage{graphicx}
\usepackage{empheq}
\usepackage{ifthen}
\usepackage{tensor}
\usepackage{paralist}
\usepackage{float}

\usepackage{slashed}
\usepackage{amsmath}
\usepackage{amsfonts}
\usepackage{amssymb}
\usepackage{mathrsfs}
\usepackage{extarrows}
\usepackage{verbatim}
\usepackage{upgreek}
\usepackage{wasysym}

\usepackage{natbib}
\usepackage{multirow}

\usepackage{makecell}
\usepackage{array}
\usepackage{boldline}
\usepackage{subcaption}

\usepackage{lettrine}
\usepackage{graphicx}
\usepackage{booktabs}

\title{A Bright Future? Prospects for Cosmological Tests of GR with Multimessenger Gravitational Wave Events}

\author[a]{Elena Colangeli,}
\author[a]{Konstantin Leyde,}
\author[a]{and Tessa Baker}

\affiliation[a]{\begin{center} \textit{Institute of Cosmology and Gravitation, University of Portsmouth, \\
Burnaby Road, Portsmouth PO1 3FX, United Kingdom}\end{center}}

\emailAdd{elena.colangeli@port.ac.uk}
\emailAdd{konstantin.leyde@port.ac.uk}
\emailAdd{tessa.baker@port.ac.uk}

\abstract{Further bright sirens -- gravitational wave events with electromagnetic counterparts -- are keenly awaited, but proving elusive. The exceptional event GW170817 had a profound impact on the landscape of viable cosmological extensions of General Relativity (GR); can we expect this kind of shift to be repeated in the next decade? In this work we will assess the potential constraints from bright sirens in the LIGO--Virgo--KAGRA O5 era and third generation detector era. We set up the statistical formalism for our constraints, and generate and analyse simulated data in the context of general scalar-tensor theories. We highlight the important role that gamma-ray burst detection has in breaking key parameter degeneracies. We find that the next ten bright sirens alone will not competitively constrain cosmological gravity, but that one year of third generation observations could confidently detect mild departures from GR, e.g. the Horndeski parameter $\alpha_{\rm M}\neq 0$ is detected at greater than $3\sigma$. This justifies investment in a broad range of methods for gravitational wave cosmology (dark sirens, bright sirens and cross-correlation with large-scale structure) to ensure tests of cosmological gravity advance in both the short-term and the long-term.}

\begin{document}
\maketitle
\flushbottom

\section{Introduction}
\label{sec:Introduction}

Gravitational waves (GWs) from compact binary coalescences offer powerful new opportunities for testing General Relativity (GR) and its alternatives.
GW generation, propagation, and polarization of these signals can all be modified in an extended gravity scenario, either coherently or independently depending upon the nature of the deviation from GR. This work will focus on deviations from GR affecting cosmological scales (see \cite{Clifton:2011jh} for a comprehensive review), where the accelerated expansion of the universe has raised questions about the completeness of our  theory of gravity. On these large scales, GW detections serve as highly complementary probes to well-established observables such as galaxy clustering, weak lensing, supernovae and the cosmic microwave background (CMB). 

A property frequently exploited in cosmological tests of GR is that gravitational waves act as standard sirens, meaning their \textit{gravitational wave luminosity distance}, $d_{\rm GW}$, can be measured directly from the amplitude of the GW signal \cite{Schutz:1986gp}. In  modified gravity theories this distance often differs\footnote{Though this feature is very widespread, counter-examples exist, e.g. see \cite{Linder:2019bqp}.} from the regular luminosity distance associated to a redshift in a Friedmann-Robertson-Walker universe, $d_{\rm L}$ (eq.~\ref{eq:dGW} below). Hence, if independent information on both redshift and the gravitational wave luminosity distance can be obtained for a GW source, the ratio $d_{\rm GW}/d_{\rm L}$ serves as a generic litmus test for departures from GR (indicated if it differs from unity).

There are two main types of GW signals that can be used for this type of analysis: dark and bright sirens. Dark sirens are events without a detected electromagnetic (EM) counterpart; to perform cosmological analyses, one associates the event to a redshift distribution via the galaxy catalogue method \cite{DES:2019ccw, Gray:2019ksv}, the spectral sirens method \cite{Taylor:2011fs, Mastrogiovanni:2021wsd, Ezquiaga:2022zkx}, or ideally both methods simultaneously \cite{Gray:2023wgj, Mastrogiovanni:2023emh}. All events with sufficiently high signal-to-noise (SNR) detected by the LIGO--Virgo--KAGRA (LVK) network can be used as dark sirens, hence we are `guaranteed' to obtain these signals. Their disadvantage is that the constraints obtained are strongly dependent on the provision of a high-completeness galaxy catalogue in the event localisation volume. Prior to public data releases from Stage IV surveys \cite{LSST:2008ijt, EUCLID:2011zbd, DESI:2016fyo}, a deep all-sky catalogue has not been available.

Bright sirens are events with an EM counterpart, most likely produced in mergers involving a neutron star. The EM counterpart can be produced in the form of a short gamma-ray burst (GRB) and a kilonova\footnote{There is additionally a GRB afterglow and X-ray radiation, though these will not be used in this work.}. The nature of these counterparts makes identification of a unique host galaxy possible, leading to a high-precision redshift measurement and hence the strongest achievable constraints on cosmology and gravity. Indeed, the detection of the binary neutron star (BNS) event GW170817 \cite{LIGOScientific:2017vwq, LIGOScientific:2017zic} provided a measurement of the Hubble parameter of $H_0 = 70^{+12.0}_{-8.0}\, \text{km s}^{-1}\text{Mpc}^{-1}$ \cite{LIGOScientific:2017adf}. However, with only one bright siren detected in nearly a decade of (admittedly non-contiguous) LVK operation, the rate of BNS mergers is now suppressed to a range between $10 \,\text{Gpc}^{-3}\text{yr}^{-1}$ and $1700 \,\text{Gpc}^{-3}\text{yr}^{-1}$ (90\% confidence interval)\cite{KAGRA:2021duu}. 

This leads us to the question: which variety of sirens -- bright or dark -- should we look to control the landscape of ideas about gravity and dark energy? Whilst dark sirens have received the greatest investment of time and energy in recent years, there is often an unspoken impression that a single additional bright siren could completely blow these constraints out of the water. In this work we evaluate to what extent this is actually possible.



Thus far we have discussed bright and dark sirens as tools to probe the distance ratio $d_{\rm GW}/d_{\rm L}$. The full reality of such an analysis is more complex. Firstly, the Hubble constant -- itself a major source of tension \cite{Verde:2019ivm, Freedman:2017yms, DiValentino:2021izs, Hu:2023jqc} -- enters the computation of luminosity distances. It must therefore be co-varied alongside any modified gravity parameters\footnote{Of course, one can turn the problem around and consider bright/dark sirens to constrain $H_0$ alone purely within the context of $\Lambda$-Cold Dark Matter ($\Lambda$CDM) cosmology. This has been covered extensively by other authors \cite{Holz:2005df, Chen:2017rfc, Palmese:2021mjm, LIGOScientific:2021aug, Mastrogiovanni:2022hil, Mukherjee:2022afz, Palmese:2023beh, Gray:2023wgj}.}. Secondly, a modified gravity theory will have additional effects beyond the GW luminosity distance which must be accounted for. One such possibility is a change to the speed of propagation of GWs, $c_{\rm T},$ a feature that can be measured in the presence of a GRB by comparing arrival times of the EM and GW signals. At low redshifts it is strongly constrained to  $-3\times10^{-15} \leq c_{\rm T}/c -1 \leq +7 \times 10^{-16}$ \cite{ LIGOScientific:2017zic} by bright siren GW170817 and its counterpart GRB170817a \cite{ LIGOScientific:2017zic}. This result follows the assumption that $c_{\rm T}$ is constant, while a non-luminal or non-constant GW propagation speed impacts the more open luminosity distance constraints discussed above.  
These degeneracies have often not been fully accounted for in other works \cite{Baker:2019gxo, Mukherjee:2020mha, Mastrogiovanni:2020gua, Mastrogiovanni:2020mvm, Mancarella:2021ecn, Leyde:2022orh, Chen:2024xkv}. In this paper we display the computations necessary for such corrections and assess their relevance.

This paper will focus on the Horndeski family of modified gravity theories (the merits of which we will cover in the next section), aiming to realistically predict the possibilities of constraining $H_0$ and beyond-GR parameters with bright sirens within the next planned LVK observing runs 
and next generation detectors, specifically the Einstein Telescope (ET). 

We present forecasts for O4/O5 in which we both constrained GR and cosmology jointly and GR alone. We will refer to dark sirens results to compare the constraining power of the two categories of events, evaluating what would be needed for bright sirens to really change the picture.  Although we will frame our quantitative forecasts in terms of Horndeski gravity, we expect our qualitative conclusions to hold more generally for tests of cosmological modified gravity.

The structure of this paper is as follows: Section~{\ref{sec:MG}} covers Horndeski theory and its impact on GW propagation. In Section~{\ref{sec:Simulation}} will cover the mock data and simulation setup used for the LVK O4/O5 case and next generation detectors, while Section~{\ref{sec:Statistical Framework}} we outline the Bayesian framework in both scenarios. In Section~{\ref{sec:Results}} we show our results and discuss our findings.


\section{Horndeski Gravity}
\label{sec:MG}

\subsection{Motivation}

In a full theory of gravity, any modifications to GW luminosity distances and GW propagation speed will be linked by new parameters appearing in the gravitational Lagrangian. Whilst modifications to the GW luminosity distance and speed can be parameterised independently in model-agnostic analyses, this yields weaker constraints and separates the two effects in an unrealistic way. 
In this work we will avoid this artificial weakening, and link these GW properties together in the framework of generic scalar--tensor theories, which add a scalar degree of freedom to the Einstein-Hilbert action. This family of theories is described by the Horndeski class, which is the most general scalar--tensor theory producing second-order field equations in four dimensions \cite{Horndeski:1974wa,Kobayashi:2019hrl}.


Horndeski gravity is an ideal testing ground for efforts to constrain modified gravity because it treads a fine balance between generality, observational viability and computational feasibility. The Lagrangian is general enough to subsume many mainstream modified gravity models, whilst compact enough to introduce a handful of additional parameters. The evolution of large-scale structure (LSS) can be computed using adapted Einstein-Boltzmann solvers on linear scales \cite{Hu:2013twa, Raveri:2014cka, Zumalacarregui:2016pph}, and simulated numerically on nonlinear scales \cite{Bose:2020wch, Wright:2022krq, Gupta:2024seu, Bose:2024qbw, Gordon:2024jaj}. 
As a result, some Horndeski parameters have been bounded -- but not fully constrained --  by EM observables \cite{Bellini:2015xja, Renk:2017rzu, Noller:2020afd, Seraille:2024beb} and the GW propagation speed bounds from GW170817 \cite{Baker:2017hug, Creminelli:2017sry, Ezquiaga:2017ekz, Kase:2018aps, LIGOScientific:2018dkp}, yielding a useful prior volume in which to pursue our work. 


These parameters modify the behaviour of perturbations, the mathematical description of which we will explore in the next section. Qualitatively, modifications to the shape of the power spectrum, the coupling of matter to gravity, and the speed of sound of perturbations renders the CMB, ISW, redshift space distortions, and baryon acoustic oscillations (BAO) particularly significant probes. Horndeski parameters bounds from analyses of these EM sources are all currently consistent with GR. In particular, $\aM$, which parametrises the coupling of matter to gravity (and will be fully introduced in the next section), has been found to be $-1.65 < \aM < +0.22$ (99.73\% limits) {\cite{Bellini:2015xja}} and $-0.62 < \aM < +1.35$ (95\% limits) {\cite{Noller:2018wyv}}. Most recently, the DESI collaboration published the most stringent result yet (for a specific parametrisation of the $\aM(z)$ function, which will be discussed in the next sections) of $\aMzero = 0.98 \pm 0.89$ (68\% confidence level) {\cite{DESI:2024yrg}}.

The coupling to gravity is also inferrable with gravitational waves, along with modifications to their propagation speed. The latter effect has been strongly constrained, as previously mentioned, to modifications of order $\sim 10^{-15}$. On the other hand, GWs have yielded less constraining results for the $\aM$ parameter, both due to the scarcity of data (hundreds of events, compared to DESI's 4.7 million data points) and their large uncertainties. Nonetheless, the best constraint on $\aMzero$ from GW dark sirens for the $\Omega_\Lambda$ parametrisation is $\aMzero = 1.5^{+2.2}_{-2.1}$ {\cite{Chen:2023wpj}} (with similar results found by {\cite{Leyde:2022orh}} and {\cite{Mancarella:2022cgn}}). 
The latest gravitational wave catalogs, GWTC-3 and GWTC-4, have expanded the available dataset significantly, but the statistical power of gravitational wave constraints remains limited compared to EM-based probes. GWTC-3, for instance, increased the number of detected events to 90 {\cite{KAGRA:2021vkt}}, while GWTC-4 is expected to further expand this dataset. Nevertheless,  data quantity and quality are limited compared to EM-based probes,  This, along with our results later on suggest that bright sirens alone are unlikely to provide conclusive constraints on $\aM$ with LVK data. .


In this section we introduce the action of Horndeski theory, one formulation of its key parameters, and quantify its effects on luminosity distances and GW propagation speed.


\subsection{Model and Parameters}
Horndeski gravity is described by the following Lagrangian \cite{Deffayet:2011gz}:

\begin{equation}
\label{eq:action}
    S = \int d^4x \sqrt{-g} \bigg[
    \sum_{i=2}^5 \mathcal{L}_i (\phi, g_{\mu\nu})\bigg] \; ,
\end{equation}
where the $\mathcal{L}_i$ are given by:
\begin{align}
    \nonumber \mathcal{L}_2 &= K(\phi, X) \;,
    \\ \nonumber \mathcal{L}_3 &= -G_3(\phi, X) \Box \phi \;,
    \\ \nonumber \mathcal{L}_4 &= G_4(\phi, X)R+G_{4X}(\phi, X) \big[(\Box \phi)^2-\nabla_\mu \nabla_\nu \phi \nabla^\mu \nabla^\nu \phi \big] \;,
    \\ \nonumber \mathcal{L}_5 &= G_5(\phi, X)G_{\mu \nu} \nabla^\mu \nabla^\nu \phi \;,
    \\ \nonumber &- \frac{1}{6}G_{5X}(\phi, X) \big[(\Box \phi)^3 + 2 \nabla_\mu \nabla^\nu \phi \nabla_\nu \nabla^\alpha \phi \nabla_\alpha \nabla^\mu \phi - 3 \nabla_\mu \nabla_\nu \phi \nabla^\mu \nabla^\nu \phi \Box \phi \big] \; .
\end{align}
The functions $K(\phi, X)$ and $G_i(\phi, X)$ ($i=3,4,5$) are arbitrary functions of the scalar field $\phi$ and its kinetic term $X = -\nabla^\mu \phi \nabla_\mu \phi /2$; the subscripts $\phi$ and $X$ indicate derivatives with respect to these quantities. One recovers the GR action when $G_4=\frac{1}{2}M^2_{\rm Pl}$ and all other $K$ and $G$ functions above vanish. Assuming a linearly perturbed FLRW metric, \cite{Bellini:2014fua} introduced a popular repackaging of the Lagrangian functions above into the four time-dependent "alpha functions". A specification of these objects is sufficient to fully describe the \textit{linear perturbative} dynamics of a Horndeski model (though see \cite{Langlois:2015cwa, Hirano:2019scf, Gleyzes:2014dya,  LISACosmologyWorkingGroup:2019mwx} for extensions of the original Horndeski framework; these are now strongly constrained on cosmological scales \cite{Crisostomi:2017pjs, Hiramatsu:2022fgn, Kase:2018iwp, Sakstein:2016ggl, Dima:2017pwp}, so we do not consider them here). Qualitatively, the alpha functions are: 
\begin{itemize}
    \item $\aM(z)$, which represents the 
    change in the effective gravitational coupling strength, $G$, connected to an effective running Planck mass (more on this below);
    \item $\aT(z)$, which describes the relative difference between the propagation speed of gravitational waves and that of light;
    \item $\alpha_{\rm K}(z)$ or `kineticity', quantifying the kinetic energy of the scalar perturbations and affecting their sound speed;
    \item $\alpha_{\rm B}(z)$ or `braiding' term, which describes the mixing of the scalar field and metric kinetic terms, causing dark energy to cluster.
\end{itemize}
In this work we will study the first two parameters, which influence tensor perturbations, and how to probe them with bright sirens. The latter two parameters affect scalar perturbations and hence are better constrained by large scale structure, see e.g. \cite{Bellini:2015xja, Renk:2017rzu, Noller:2018wyv}. To formulate these one must first define the following quantity, denoted by $M_*$, which is related to the Lagrangian operators in Eq.\ref{eq:action} by \cite{Bellini:2014fua}:
\begin{equation}
\label{eq:M_*}
    M^2_*(\phi, X, H) = 2(G_4 - 2XG_{4X}+XG_{5\phi}-\dot{\phi}HXG_{5X}) \; ,
\end{equation}
where dots represent derivatives with respect to coordinate time. This is the quantity that `acts' as a Planck mass in this theory, but unlike in GR, it does not have to be constant over time. We can capture this time evolution explicitly though the form of $\aM$ (where we retain $z$-dependence on the LHS, but indicate on the RHS that a derivative with respect to the scale factor is often the most useful):
\begin{align}
\label{eq:alphaM definition}
    \aM(z) &= \frac{\dd \ln{M_*^2}}{\dd \ln{a}} \;.
\end{align}
The tensor speed parameter $\aT$ quantifies deviations in the speed of tensor perturbations $c_{\rm T}$ as
\begin{align}
\label{eq:alphaT definition}
c^2_{\rm T} &= c^2(1+\aT(z)) \; .
\end{align}
The parameter $\aT$ can also be expressed in terms of the $G_i$ functions above: 
\begin{equation}
\label{eq:M_*aT}
    M^2_*\aT=2X[2G_{4X}-2G_{5\phi}-(\ddot{\phi}-\dot{\phi} H)G_{5X}] \; .
\end{equation}
It is important to note that $\alpha_{\rm B}$ does appear in the modified Friedmann equations in Horndeski gravity, however one can choose this function freely to reproduce any expansion history. The cosmological background can then be fully described by a function of time such as $H(z)$ or $w(z)$ (the dark energy equation of state). Instead these $\alpha$-parameters only affect perturbations, hence we probe $\aT$ most directly through gravitational waves, while $\alpha_{\rm B}$ and $\alpha_{\rm K}$ are examined looking at CMB and LSS data, and $\aM$ features in both GWs and LSS \cite{Bellini:2015xja, Seraille:2024beb}. To isolate the effects of Horndeski alpha parameters, the cosmological background in this work is taken to be $\Lambda$CDM with $\Omega_{\rm m} = 0.3065$, consistent with Planck 2018 constraints \cite{Planck:2018vyg}. 

GWs propagate differently in Horndeski gravity than in GR. This is obtained by varying the action above w.r.t the metric perturbation $h_{\mu\nu}$ and extracting the equations of motion for the tensorial degree of freedom. Evaluating the resulting GW propagation equation on an FRW metric gives \cite{Riazuelo:2000fc, Saltas:2014dha, Belgacem:2018lbp}: 
\begin{align}
\label{modified propagation}
    h''_{ij}+[2+\aM(z)]\mathcal{H}h'_{ij}+c^2_{\rm T}k^2 h_{ij} =0\;.
\end{align}
Here $\aM$ acts as a friction term affecting the amplitude of GWs while $\aT$, contained in $c_{\rm T}$, regulates the speed at which they propagate. These two quantities can be constrained from two main observables: distances and time delay in arrival of GW and EM signals. The model used in this paper assumes that production of GWs is the same as in GR, with the modified gravity effects only affecting propagation. 

The $\aT$ and $\aM$ functions are not independent of each other in a model where $c_{\rm T}$ is not constant \cite{Kennedy:2017sof}. However, 
the effect of $\aT$ on $\aM$ is negligible as explored in Appendix~\ref{app: aT and aM}, allowing us to analyse the two independently. 

The $\aM$ term arises due to an evolving gravitational self interaction $G_{\rm gw}(t)$ which differs from the standard matter coupling in GR, $G_{\rm N}$. Note that these two are two separate quantities which are equal in GR. While one can measure $G_{\rm N}$ locally, standard sirens are the best way to probe $G_{\rm gw}(t)$ as shown by \cite{Wolf:2019hun}.


\subsection{Parametrisation of \texorpdfstring{$\alpha_i(z)$}{Lg}}
\label{subsec:param}

In order to constrain $\aM(z)$ and $\aT(z)$, we have to make an ansatz for their redshift-dependence\footnote{This can be computed explicitly in a given modified gravity model, but here we wish to remain more agnostic -- hence we use a motivated phenomenological ansatz.}. The most common ansatzes are:
\begin{align}
\label{eq:parametrisations}
    \alpha_i(z) = \alpha_{i0}\Omega_\Lambda(z)/\Omega_{\Lambda 0}, \quad\quad\; \; \; \alpha_i(z) = \alpha_{i0}a, \quad\quad\; \; \; \alpha_i(z) = \alpha_{i0}a^p
\end{align}
 as seen in \cite{Baker:2020apq, Lagos:2019kds, Seraille:2024beb}. These ansatzes all modulate the effects of modified gravity to 
 be negligible at high redshifts ($\alpha_i(z)\rightarrow 0$), as we do not wish to modify matter and radiation dominated eras. 
 At low redshifts the $\alpha_i$ approach unity.
 
 Although the $\alpha_i$ are largest at low redshifts, their impact on GW luminosity distances and arrival times is an integrated quantity over their propagation distance (see \ref{eq:dGW} and \ref{eq:SoG} in the later sections). For this reason, differentiating between the three ansatzes above will yield less significant changes at LIGO--Virgo--KAGRA ranges, and starts making a more marked at higher redshifts. This property can be exploited with ET, which is predicted to detect neutron stars coalescences up to redshift $\sim 2-3$ \cite{Maggiore:2019uih, Reitze:2019iox}, as will be discussed later. Since parametrisation choices could impact results, ideally multiple parametrisations should be tested when using real data to assess any differences, particularly at ET range. However, differences are not expected to exceed order unity for our ET forecast (and they are negligible for the LVK case), hence we use the widely adopted $\Omega_\Lambda$ parametrisation \cite{Bellini:2014fua, Bellini:2015xja, Alonso:2016suf} to facilitate comparison with current constraints. Studies are underway to constrain these functions in a non-parametric manner, e.g. \cite{Belgacem:2019zzu} and \cite{Afroz:2023ndy, Afroz:2024joi, Afroz:2024oui}, though they require much larger amounts of data to reach similar constraining power as a parametric method.

As mentioned above, current GW constraints on $\aMzero$ come from dark sirens. For the $\Omega_\Lambda$ parametrisation we will compare our findings to $\aMzero = 1.5^{+2.2}_{-2.1}$ {\cite{Chen:2023wpj}} and the DESI result of $\aMzero = 0.98 \pm 0.89$ (68\% confidence level) {\cite{DESI:2024yrg}}, which was obtained marginalising over other $\alpha_{\rm B}$.
As for the tensor speed excess term,  $|\aT| \lesssim  10^{-15}$ \cite{LIGOScientific:2017zic} as previously mentioned. Despite this constraint having been obtained under the assumption of a non-evolving $\aT$, it can be used as a prior in our case since $\aT(z)$ can be approximated to a constant at very low redshifts (as we will see in a moment). GR is recovered when these functions are set to zero.

\subsection{Distances in Horndeski}
\label{ssec:Distances in MG}

The amplitude of $h$ (obtained by solving eq. \ref{modified propagation}) leads to the relation between EM luminosity distances and GW distances in Horndeski gravity, which includes corrections arising from an evolving GW speed \cite{Mastrogiovanni:2020gua, LISACosmologyWorkingGroup:2022wjo}. The impact of $\aT$ on luminosity distance is negligible as errors on $d_{\rm GW}$ are comparatively large, and hence this term can be dropped, leading to \cite{Belgacem:2017ihm, Belgacem:2018lbp}: 
\begin{align}
\label{eq:dGW}
    d_{\rm GW} = d_{\rm L} \text{ exp} {\bigg\{\int_0^z\frac{\aM(z') \dd z'}{2(1+z')}\bigg\}}
\end{align}
where the EM luminosity distance is defined as usual:
\begin{align}
    d_{\rm L} = \frac{(1+z)}{H_0} \int_0^z \frac{ \dd z'}{E(z')} \text{ .}
\end{align}
In this expression $E(z) = H(z)/H_0$.
%
As hinted above, we see in eq.~\ref{eq:dGW} that the modified gravity effect is integrated over redshift, which implies that events that are further away will produce a more informative posterior on $\aMzero$.

\begin{figure}[h]
    \centering
    \includegraphics[width=0.6\linewidth]{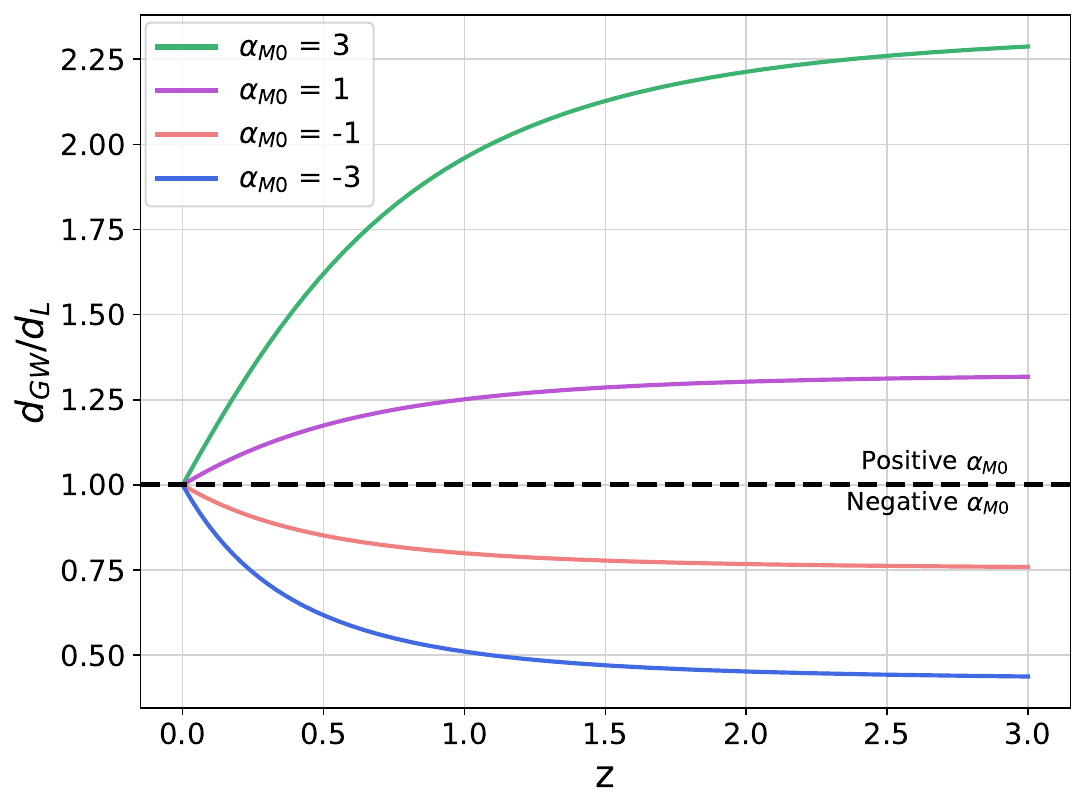}
    \caption{Ratio of GW and luminosity distances as a function of redshift for the  $\Omega_\Lambda$ parametrisation~\ref{eq:parametrisations} of $\aM(z)$. Each line corresponds to a different value of $\aMzero$.}
    \label{fig:dgwbydL_2}
\end{figure}

Figure~\ref{fig:dgwbydL_2} shows the impact that different values of $\aMzero$ have on the ratio of GW and EM luminosity distances as a function of redshift. As expected, higher absolute values of $\alpha_{M0}$ produce more pronounced deviations, with $d_{\rm GW} > d_{\rm L}$ for positive values and the opposite for negative values. In the low-redshift limit ($z \rightarrow 0$), the results converge to those predicted by GR, as the range of the integral in eq.~\ref{eq:dGW} vanishes, while deviations from GR grow more significant at higher redshifts. Consequently, distant events will be more sensitive to potential departures from GR, providing tighter constraints.

\subsection{GW Propagation Speed}
\label{ssec:Speed of Gravity}

One can show that the $\aT$ function governing the speed of gravity (eq.~\ref{eq:alphaT definition}) affects the difference in arrival times of the GRB and GW signals. The computation is performed by equating the path of a photon and of a GW emitted with a time delay $\DTe = t_{\rm e, EM} - t_{\rm e, GW}$. The arrival time delay $\DTa = t_{\rm a, EM} - t_{\rm a, GW}$ can then be expressed as a function of $\DTe$, redshift $z$, and the constant coefficient $\aTzero$:
\begin{align}
\label{eq:SoG}
     \DTa = \frac{\aTzero}{2}\bigg[\int_0^z \frac{1}{H_0 \times E(z')}\frac{\Omega_\Lambda(z')}{\Omega_{\Lambda0}} \,\dd z'\bigg] + (1+z) \DTe \; .
\end{align}
This expression is for the $\Omega_\Lambda$ ansatz selected in section~\ref{subsec:param}; a detailed derivation is provided in Appendix~\ref{app:SoG Derivation}. The typical form of $\DTa$ which was used in the analysis of GW170817 \cite{LIGOScientific:2017zic} is different from the above as it represents a universe with a constant speed $c_{\rm T}$. 
This expression has two terms: the term inside the brackets, which acts as an effective comoving distance -- analogously to the comoving distance appearing in the constant $c_{\rm T}$ case \cite{Romano:2023bge} -- and the $(1+z) \DTe$. For a range of redshifts up to a threshold, the effective comoving distance dominates the above expression. Once the threshold is reached, the second term begins dominating. Numerically exploring the relation between these two contributions we find that this switch, which depends on the cosmology and emission time delay of each event, is generally $z \sim 2$ for our parameter space. This shift in the dominant term does not affect LVK range events as they are distributed up to $z \sim 0.06$. It may affect events in the ET range though not in our work, as we impose a selection cut on distance due to the GRB detection horizon (further details in Section~\ref{ssec:next gen setup}) which restricts us to redshifts below $z \approx 0.5$.

Given that the redshift of a bright siren event can be determined from observations of the kilonova, and we will marginalise over the emission time delay, the remaining observable required for this analysis is the arrival time delay. This can only be obtained if the GW event has an associated GRB detection. The production mechanism of short GRBs from mergers is not fully understood, however detection depends on the jet opening angle and the inclination of the binary \cite{Lazzati:2017zsj, Lamb:2017ych}. In this study we assume GRB detection occurs for binary systems with inclinations less than $20^\circ$ (or greater than $160^\circ$ due to the bipolar nature of the jets) as the average jet opening angle has been found to be $\sim 10^\circ$ \cite{Escorial:2022nvp, Jin:2017hle, Ghirlanda:2016ijf}. 

\section{Mock Data}
\label{sec:Simulation}

\subsection{LIGO--Virgo--KAGRA Network}
In this section we will describe the mock data generation process for the LVK scenario, outlining the Bayesian formalism in the next. The methods employed for the ET case, which differ from those used for LVK, will be discussed in a later section.

The number of BNS mergers expected to be detected in O4 and O5 is low, in fact the upper bound of the aforementioned rate ($1700 \,\text{Gpc}^{-3}\text{yr}^{-1}$) implies only 30 events per year at current sensitivity. Here we consider a population of $10$ events observed in one year which corresponds to a rate of approximately 500 events $\text{Gpc}^{-3}\text{yr}^{-1}$, with one of these events having an associated GRB due to its inclination of $\iota = 6.3^\circ$, which was sampled from a $p(\iota) \propto \sin(\iota)$ inclination prior. As mentioned in the previous section, we impose an observational cutoff for GRBs at inclinations of $\iota < 20^\circ$ (face-on) or $\iota > 160^\circ$ (face-off) since jets are bipolar. Employing a uniform in $\sin(\iota)$ prior on inclination results in $\sim 6\%$ of events having a GRB, hence we cosider only one event to have an associated observed GRB. 


All events have redshift information since kilonovae are isotropic, allowing for precise identification of the host galaxy. GW170817 is not analysed, however the results obtained from the event are used to inform our priors on $H_0$ and $\aTzero$.

Our simulations rely on three datasets: distance posteriors from GW detections, time delays in the arrival of GW and GRB signals, and redshifts from the bright EM counterpart. Distance posteriors were generated using the \texttt{bilby} package \cite{bilby_paper} and the \texttt{bilby\_pipe} tool \cite{bilby_pipe_paper} with projected O4 detector sensitivity (see Appendix~\ref{app:bilby} for the specifics). The model assumed zero noise and fixed phase, time of coalescence, tidal deformability parameters, spins, and sky location. The first four parameters do not significantly impact the error on the distance posterior, while sky location can be fixed due to the precise localisation of the kilonova. For simplicity, the sky position for all events is chosen to be optimal for the LIGO Hanford detector. The parameters that are inferred alongside the distance are masses and inclination, as discussed in the previous sections. Events are assumed to be distributed uniformly in comoving volume, with masses sampled according to a uniform distribution in source frame between $1$ and $3 M_\odot$ \footnote{Limits are consistent with the literature \cite{Landry:2021hvl, Chu:2024rwt}, a uniform prior is chosen for simplicity but it does not affects results as we marginalise over masses and remain consistent when computing selection effects (see next section).}, and inclinations sampled from a prior that is uniform in sine.

In the case of the event with corresponding GRB, a prior for the emission time delay needs to be specified. We adopt a flat distribution between $-1$ and $10$ seconds, where these bounds are informed by jet physics, as detailed in \cite{Zhang:2019ioc}. A more restrictive prior -- assuming little to no time delay between the two signals -- can lead to biases in the estimation of $\aTzero$ if this is not the case in nature. On the other hand, the arrival time delay measurement is known with high precision and is modelled as a narrow Gaussian around the true value, assuming an $1\sigma$ error of $0.1$ s. This is consistent with the observation of GW170817, which had time delay of $1.74 \text{s} \pm 0.05$ s \cite{ LIGOScientific:2017zic}. All events are simulated according to a cosmology with $H_0=70\, \text{km s}^{-1}\text{Mpc}^{-1}$ and the $\Omega_\Lambda$ parametrisation of the $\alpha$ parameters with $\aMzero = 1$ and $\aTzero = 2.699 \times 10^{-16}$, both within the current inferred bounds\footnote{$\aTzero$ was chosen using the above cosmology, the GW170817 data, and drawing an emission time delay value from the $\aTzero$ prior.}.

In order to avoid biases, we use selection effects (which we will formally define later in~\ref{eq:selection_eff}) that are consistent with the approximations made, depending only on distance, masses, and inclination. Note that selection is determined solely by the detection horizon of the interferometers and is not influenced by GRB data, i.e. we analyse all data, irrespective of whether it has a counterpart or not. The network signal-to-noise ratio threshold for detection is set to $12$ in the simulations.

Two scenarios are considered: one in which we both test GR and infer cosmology and a more `optimistic' one where we only test GR. In the first scenario (which we'll call `GR + Cosmology') uninformative priors are used for $H_0$, $\aMzero$, and $\aTzero$, with $\Delta t_{\rm e}$ following the aforementioned flat prior (between -1 and 10 seconds). The second scenario is based on the (hopeful) prediction that the Hubble tension will be resolved by the end of O5 (early 2030s), changing the $H_0$ prior to a Gaussian centred at the true value with errors comparable to Planck \cite{Planck:2018vyg}. The priors for $\aMzero$, $\aTzero$ and $\DTe$ remain unchanged. A full summary of the priors used in both scenarios can be found in Table~\ref{tab:priors_H_alphas}. 

\begin{table}[H]
\renewcommand{\arraystretch}{1.2}
\begin{center}
\caption{Table of priors. Note that the true emission time delay is for the one event with GRB only.}
\label{tab:priors_H_alphas}
\begin{tabular}{ !{\vrule width 0.5mm} c | c | c !{\vrule width 0.5mm} } 
\hlineB{3}
Parameter & True & Prior \\
\hline
$H_0$ & $70\, \text{km s}^{-1}\text{Mpc}^{-1}$ & 
\begin{tabular}{@{}c@{}} $\mathcal{U}[50, 90]\, \text{km s}^{-1}\text{Mpc}^{-1}$ (GR + Cosmology) \\ $\mathcal{N}(70, 1)\, \text{km s}^{-1}\text{Mpc}^{-1}$ (Testing GR only) \end{tabular} \\
$\alpha_{\rm M0}$ & $1$ & $\mathcal{U}[-15, 15]$ \\
$\alpha_{\rm T0}$ & $2.699 \times 10^{-16}$ & $\mathcal{U}[-500, 200] \times 10^{-16}$ \\
$\Delta t_{\rm e}$ & $2.1$ s & $\mathcal{U}[-1, 10]$ s \\
\hlineB{3}
\end{tabular}
\end{center}
\end{table}

\subsection{Einstein Telescope}
\label{ssec:next gen setup}
In addition to simulating mock bright sirens for the LVK O4/O5 era, we also perform simulations for the Einstein Telescope. Due to the large number of events considered, we simplify some details of the simulated data; given the large number of expected events, the impact of an individual event's properties on the overall result is reduced. 
As previously mentioned we expect around $6\%$ of events to have inclinations that allow for a GRB to be observed. Keeping this in mind, and allowing a fraction to go unobserved due to low luminosity (more on this later) we simulate 150 events with associated GRBs, which is a plausible estimate as forecasts indicate that ET will see up to $6 \times 10^4$ BNS events per year\footnote{Einstein Telescope: Science Case, Design Study and Feasibility Report: \url{https://www.et-gw.eu/index.php/relevant-et-documents}}. We calculate the detection horizon $d^{\rm thr}_{\rm GW}$ for an optimally placed face-on source with detector frame masses of $1.4-1.4 M_\odot$, using the ET PSD available in the \texttt{pyCBC} package \cite{pycbc} for an SNR of $12$. 
Events are once again sampled from a prior that is uniform in comoving volume. The errors in distance are proportional to the true luminosity distance, under the assumptions that the observed distances follow a Gaussian distribution centred around the true value:
\begin{align}
    p(d_{\rm GW}^{\rm obs}|d_{\rm GW}^{\rm true}) = \frac{1}{\sqrt{2 \pi} \sigma} \text{exp} \bigg\{-\frac{(d_{\rm GW}^{\rm obs}-d_{\rm GW}^{\rm true})^2}{2 \sigma^2} \bigg\} \; .
\end{align}
Here $\sigma = A \times d_{\rm GW}^{\rm true}$ where $A$ represents the fractional error. The value of $A$ used for all simulated events is $7 \%$ which was computed with \texttt{bilby} for an event at a distance of 2000 Mpc, integrating over the restricted inclination range without knowing whether the system is face-on or face-off (integrating over $0^\circ$ to $20^\circ$ and $160^\circ$ to $180^\circ$). Though not all events will have distances distributed according to the above Gaussian and have the same fractional error, 7\% is a conservative estimate leading to a conservative result. We also compute the posterior for the same event assuming no GRB is detected, meaning we integrate over the whole inclination prior to later compare analyses, which results in $A = 20\%$.
\begin{figure}[h]
    \centering
    \includegraphics[width=0.65\linewidth]{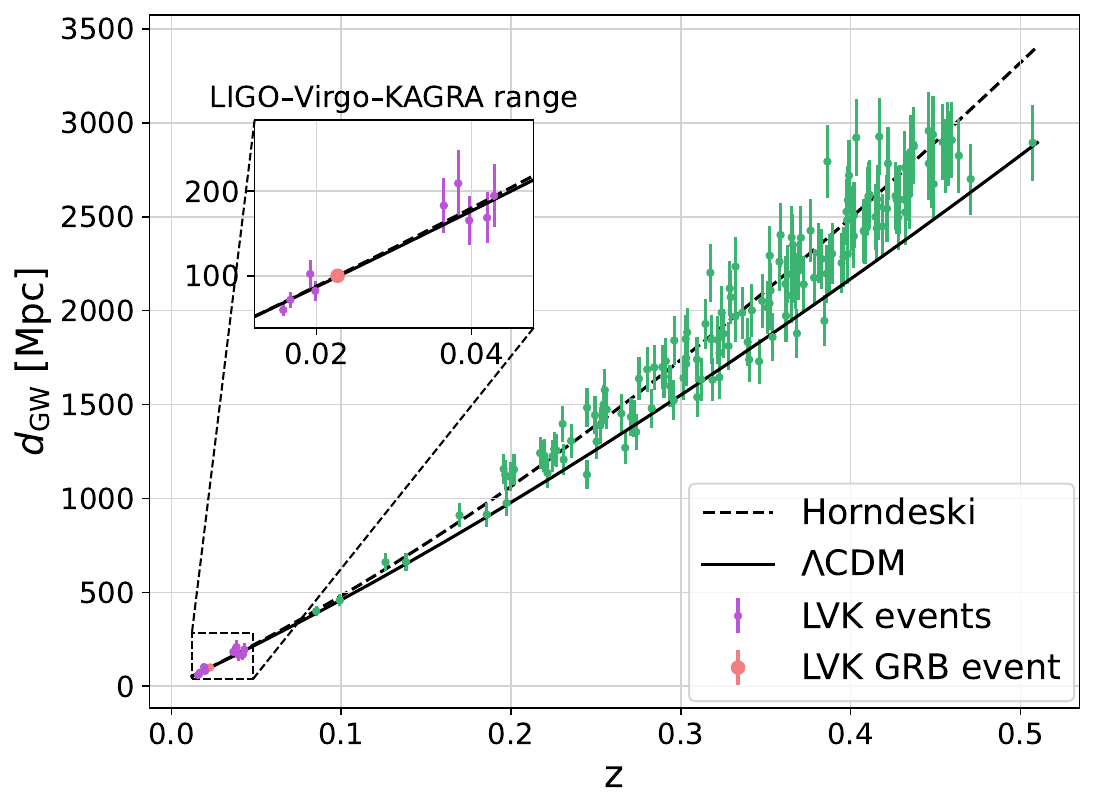}
    \caption{Plot illustrating the mock data for both LIGO--Virgo--KAGRA and ET scenarios, all data points are centred at the observed values with corresponding errors in distances on the y--axis. The green points are the 150 ET events (all with GRBs). The solid line represents $d_{\rm GW}$ as a function of redshift in the chosen fiducial cosmology with $\aMzero=1$ and $H_0=70\, \text{km s}^{-1}\text{Mpc}^{-1}$, while the dashed line corresponds to the same quantity in GR.}
    \label{fig:events}
\end{figure}
Figure~\ref{fig:events} depicts the mock data used for both LVK and ET simulations, with corresponding distance errors.

ET will be capable of detecting events at larger redshifts, however GRBs might not be detectable with Fermi (or Swift) at such distances. Though more advanced GRB detectors may be available by the time ET is functional, we model current ones so as to not overestimate the amount of detectable multimessenger events. Fermi detects GRBs if the received flux in its energy band surpasses the threshold of 3 (4 in the case of Swift) $\times 10^{-8} \text{erg}\text{s}^{-1}\text{cm}^{-2}$ \cite{Bhattacharjee:2024wyz}, where flux $F$ is defined as
\begin{align}
\label{eq:flux}
    F = \frac{L}{4\pi [d_{\rm L}(z, H_0)]^2} \;.
\end{align}
$L$ is intrinsic luminosity of the GRB, which is normally defined as a function of the jet aperture angle and inclination of the binary. Fixing these in our simulation allows us to choose a simpler model for this parameter: we draw an $L$ for each simulated GRB from a normal distribution centred around $L=5\times10^{49}\text{erg s}^{-1}$ with a width of $\sigma_{\rm L} = 0.5\times10^{49}\text{erg s}^{-1}$ (or a 10\% error). Once again we wish to remain conservative on the abundance of GRBs so we choose the mean for $L$ from the fainter end of the distribution in \cite{Ghirlanda:2016ijf} and  \cite{Wanderman:2014eza} rather than the peak luminosity they find. We use $L$ and the redshifts to compute the corresponding flux for the fiducial cosmology with $H_0 = 70 \text{km s}^{-1}\text{Mpc}^{-1}$, selecting events that are above the Fermi threshold. The number of detected GRB events reaches almost zero at $d_{\rm L} \sim 3000~$Mpc, similarly to \cite{Iampieri:2024dul}.

\section{Statistical Framework}
\label{sec:Statistical Framework}

\subsection{LIGO--Virgo--KAGRA Network}

The joint posterior for $\aMzero, \aTzero$, and $H_0$ for an event with GW distance data $x_{\rm GW}$, redshift data $x_{\rm EM}$ and possible GRB data $x_{\rm GRB}$ is given by:
\begin{align}
\label{eq:posterior short version}
    p(\aMzero, \aTzero, H_0 | x_{\rm GW}, x_{\rm EM}, x_{\rm GRB}) &= \frac{p(x_{\rm GW}, x_{\rm EM}, x_{\rm GRB} | \aMzero, \aTzero, H_0)p(\aMzero, \aTzero, H_0)}{p( x_{\rm GW}, x_{\rm EM}, x_{\rm GRB})} \; .
\end{align}
The form of the likelihood differs between events with associated GRB detection and ones without. We use the notation $\mathcal{L}_{\rm GRB}$ to represent the likelihood for events with detected GRBs, while $\mathcal{L}_{\rm \overline{GRB}}$ denotes the likelihood for events with no GRB detection. For an event with no detected GRB the $x_{\rm GRB}$ term is not present, and $\aTzero$ cannot be constrained. Removing these, we can expand the likelihood $\mathcal{L}_{\rm \overline{GRB}} = p(x_{\rm GW}, x_{\rm EM} | \aMzero, H_0)$ over relevant parameters of the binary as follows:
\begin{align}
\label{eq:noGRB posterior}
    \nonumber \mathcal{L}_{\rm \overline{GRB}} \propto &\int_{z = 0}^{z_{\rm max}} \int_{d_{\rm GW}=0}^{d_{\rm GW}(z_{\rm max}, \aMzero, H_0)} \int_{\iota=0}^{180^\circ} \int_{m_{1,2} = 1M_\odot}^{m_{1,2} = 3M_\odot} p(x_{\rm GW}, x_{\rm EM} |  z', d'_{\rm GW}, m'_1, m'_2, \iota', \aMzero, H_0)
    \\& p( z', d'_{\rm GW}, m'_1, m'_2, \iota'|\aMzero, H_0) \,\dd z \,\dd [d_{\rm GW}]\,\dd \iota' \,\dd m'_1 \,\dd m'_2 \; ,
\end{align}
while if a GRB is present, the likelihood $\mathcal{L}_{\rm GRB} = p(x_{\rm GW}, x_{\rm EM}, x_{\rm GRB} | \aMzero, \aTzero, H_0)$ is:
\begin{align}
\label{eq:GRB posterior}
    \mathcal{L}_{\rm GRB} \propto \nonumber& \int_{z = 0}^{z_{\rm max}} \int_{d_{\rm GW}=0}^{d_{\rm GW}(z_{\rm max}, \aMzero, H_0)} \int_{\iota=0}^{180^\circ} \int_{m_{1,2}= 1M_\odot}^{m_{1,2} = 3M_\odot} \int_{\Delta t_{\rm e} = -1s}^{\Delta t_{\rm e} = 10s} \int_{\Delta t_{\rm a}(d_{\rm GW}=0, \Delta t_{\rm e}=-1, \aT)}^{\Delta t_{\rm a}(d_{\rm GW}^{\rm max}, \Delta t_{\rm e}=10, \aTzero)} 
    \\ \nonumber &p(x_{\rm GW}, x_{\rm EM}, x_{\rm GRB} |  z', d'_{\rm GW}, m'_1, m'_2, \Delta t'_{\rm a}, \Delta t'_{\rm e}, \iota', \aMzero, \aTzero, H_0)
    \\& p( z', d'_{\rm GW}, m'_1, m'_2 \Delta t'_{\rm a}, \Delta t'_{\rm e}, \iota'|\aMzero, \aTzero, H_0) \,\dd z' \,\dd [d'_{\rm GW}]\,\dd \iota' \,\dd [\Delta t'_{\rm e}] \,\dd [\Delta t'_{\rm a}] \,\dd m'_1 \,\dd m'_2 \; .
\end{align}
In the above $z$ is redshift, $d_{\rm GW}$ is the GW luminosity distance given by eq.~\ref{eq:dGW}, $m_{1,2}$ are the masses of the neutron stars in source frame, $\iota$ is the inclination of the binary system, and $\Delta t_{\rm e}$ and $\Delta t_{\rm a}$ are emission and arrival time delay (between GW and GRB signals) respectively, as seen in eq.~\ref{eq:SoG}. The integration bounds for these variables are as described in the previous section, appearing explicitly in eq.~\ref{eq:noGRB posterior} and~\ref{eq:GRB posterior}.

GW, GRB, and redshift data realisations are independent of each other, hence the first term in both integrals above can be split accordingly, with masses only appearing in the $x_{\rm GW}$ term as they modify the GW data only. In the case with the GRB detection, the probability of $x_{\rm GRB}$ can be written as $p(x_{\rm GRB}|  z, d_{\rm GW}, \DTa, \DTe, \iota, \aMzero, \aTzero, H_0) = p(x_{\Delta t}|z, \DTa, \DTe, \aTzero, H_0)p(det_{\rm GRB}|\iota)$, where $x_{\Delta t}$ represents the observed arrival time delay and $det_{\rm GRB}$ indicates the detection of the GRB itself. We drop GW distance dependency as it does not impact the detection of a GRB at LVK range (since the GW horizon is much smaller than GRB detectors') or the time delay. Detection is then only dependent on inclination, as previously mentioned. Under the assumptions adopted in this analysis, $p(det_{\rm GRB}|\iota)$ imposes a constraint on the inclination prior, limiting the range to $[0^\circ, 20^\circ]$ (or $[160^\circ, 180^\circ]$ for face-away events). This restriction breaks the inclination--distance degeneracy producing a tighter constraint for GW distance.

\noindent The second term inside the integral can be split into $p(z, d_{\rm GW}, m_{1, 2}, \DTa, \DTe, \iota|\aMzero, \aTzero, H_0) = p(d_{\rm GW}|z, \aMzero, H_0)p(\DTa|z, \DTe, \aTzero, H_0)p(z|H_0)p(m_{1, 2})$. The gravitational wave data $x_{\rm GW}$ depends on the true distance $\hat{d}_{\rm GW}$, which in turn is a function of $z$ and the cosmological parameters as defined in equation~\ref{eq:dGW}. Specifically, $p(d_{\rm GW}|z, \aMzero, H_0)=\delta(d_{\rm GW}-\hat{d}_{\rm GW}(z, \aMzero, H_0))$. Similarly, $\Delta t_{\rm a}$ is reduced to $p(\Delta t_{\rm a} | z, \Delta t_{\rm e}, H_0, \aTzero) = \delta(\Delta t_{\rm a} - \hat{\Delta t}_{\rm a}(z, H_0, \Delta t_{\rm e}, \aTzero))$, according to the relation in equation~\ref{eq:SoG}. Spectroscopic redshifts are known with uncertainties that are significantly smaller than the uncertainty in distance, allowing the $x_{\rm EM}$ term to be approximated to a $\delta$-function, $p(x_{\rm EM}|z)=\delta(z-z_{\rm obs})$. These redshifts are also well below the chosen maximum redshift, here $z_{\rm max} = 2$.\footnote{This value is equivalent to considering the whole universe in our simulation as LVK has a horizon much smaller than $z=2$, however choosing this value is computationally advantageous compared to larger ones.} Priors for the binary component masses are defined in source frame (these are denoted by $m^{\rm s}$), while the $x_{\rm GW}$ contains information about them in detector frame ($m^{\rm det}$), hence the Jacobian $(1+z_{\rm obs})^2$ will appear in our final expression.

The complete posterior distribution for the scenario with ten bright sirens (only one with associated GRB) is:
\begin{align}
\label{eq:full posterior}
      p(\aMzero, \aTzero, H_0|x_{\rm GW}, x_{\rm EM}, x_{\rm GRB}) \propto \bigg[ \prod_{i = 1}^{n_{\rm \overline{GRB}}} \mathcal{L}_{\overline{\text{GRB}, i}} \bigg] \times \bigg[ \prod_{j = 1}^{n_{\rm GRB}} \mathcal{L}_{\text{GRB}, j} \bigg]\times \frac{p(\aMzero) p(H_0) p(\aTzero)}{{[\beta(H_0, \aMzero)]^{({n_{\rm \overline{GRB}}}+{n_{\rm GRB}})}}} \; ,
\end{align}
where ${n_{\rm \overline{GRB}}}$ is the number of events without a GRB, while ${n_{\rm GRB}}$ is the number of events with a GRB (in our case ${n_{\rm \overline{GRB}}} = 9$ and ${n_{\rm GRB}}= 1$). $\mathcal{L}_{\rm \overline{GRB}}$ represents the likelihood of the events with no GRB, while $\mathcal{L}_{\rm GRB}$ corresponds to the event with a GRB as defined above. Explicitly, these are:
\begin{align}
    \mathcal{L}_{\rm \overline{GRB}} \nonumber = &\int_{\iota = 0}^{180^\circ} \int_{m_{1,2}^{\rm det}(z_{\rm obs}, m^{\rm s}=1M_\odot)}^{m_{1,2}^{\rm det}(z_{\rm obs}, m^{\rm s}=3M_\odot)} p(x_{\rm GW}|\hat{d}_{\rm GW}(z_{\rm obs}, \aMzero, H_0), {m'}^{\rm det}_1, {m'}^{\rm det}_2, \iota')
    \\ & p(m^{\rm s}_1, m^{\rm s}_2) p(z_{\rm obs}|H_0) p(\iota') 
    \frac{1}{(1+z_{\rm obs})^2} \,\dd{m'}^{\rm det}_1 \,\dd{m'}^{\rm det}_2 \,\dd  \iota'  
    \\ \nonumber
    \\ \nonumber \mathcal{L}_{\rm GRB} = & 2 \int_{\iota = 0}^{20^\circ} \int_{\Delta t_{\rm e} = -1s}^{\Delta t'_{\rm e} = 10s} \int_{m_{1,2}^{\rm det}(z_{\rm obs}, m^{\rm s}=1M_\odot)}^{m_{1,2}^{\rm det}(z_{\rm obs}, m^{\rm s}=3M_\odot)} p(x_{\rm GW}| \hat{d}_{\rm GW}(z_{\rm obs}, \aMzero, H_0), {m'}^{\rm det}_1, {m'}^{\rm det}_2, \iota') 
    \\ \label{eq:GRBpost}\nonumber &  p(x_{\Delta t}| \hat{\Delta} t'_{\rm a}(z_{\rm obs}, H_0, \Delta t'_{\rm e}, \aTzero))
    p(z_{\rm obs}|H_0)
    p(\iota'| det_{\rm GRB})p(\Delta t'_{\rm e}) p(m^{\rm s}_1, m^{\rm s}_2)
    \\ & \frac{1}{(1+z_{\rm obs})^2} \,\dd \iota' \,\dd [\Delta t'_{\rm e}] \, \dd{m'}^{\rm det}_1 \,\dd{m'}^{\rm det}_2\; .
\end{align}
The factor of $2$ outside the integral accounts for the consideration of both face-on and face-off systems. In the above $\beta(H_0, \aMzero)$ represents the selection effects. This term normalises the likelihood by integrating the likelihood numerator over all detectable events. It must be accounted for in cosmological analyses to avoid biases arising from the fact that not all events are observed. The methodology for deriving this term is discussed in \cite{Gray:2023wgj} and follows the more general description presented by \cite{Mandel:2018mve}. In the scenario presented in this paper, the specific form of $\beta(H_0, \aMzero)$ is:
\begin{align}
\label{eq:selection_eff}
     \nonumber \beta(H_0, &\aMzero) = \int p(det_{\rm GW}|z, m_1, m_2, \iota, H_0, \aMzero)p(m_1, m_2)p(\iota)p(z|H_0) \,\dd z \,\dd \iota \,\dd m_1 \,\dd m_2
    \\& \approx \frac{1}{N_{\rm tot}} \sum_{i=1}^{N_{\rm det}} \frac{p(z_i(d_{\text{GW},i},\aMzero, H_0))p(m^{\rm s}_{1,i}(m^{\rm det}_{1,i}, z), m^{\rm s}_{2,i}(m^{\rm det}_{2,i}, z)| H_0)}{p_{\rm inj}(d_{\text{GW},i}, m^{\rm s}_{1,i}, m^{\rm s}_{2,i}, \iota_i)(1+z_i)^2|\frac{\partial d_{\rm GW}}{\partial z}|_{z=z_i}} \;.
\end{align}
This is a sum over $N_{\rm det}$ detected injected signals out of $N_{\rm tot}$ total events, whose detectability depends on masses, inclination, and redshift. In this expression $det_{\rm GW}$ denotes the probability of detecting the GW signal. Since the redshift horizon for O4 and O5 is relatively small, redshift selection effects do not appear in the expression due to the larger capabilities of EM detectors, i.e. every event has a corresponding redshift \cite{LSST:2008ijt, EUCLID:2011zbd, DESI:2016fyo, SDSS:2023tbz}. Injections are generated according to a set of $p_{\rm inj}$ priors in detector frame, introducing a factor of $(1+z)^2$ in the selection effects when transforming between source and detector frame masses (as in the likelihood). The limits of integration of the denominator remain the same as previously mentioned. This translates to the same bounds being used to form the injection priors as we need to cover the same region of parameter space. 

Note that we perform simplified analyses for a year of observations with a 100\% duty cycle for GW detectors, assuming concurrent detection of GRBs and redshifts to be possible for all detected GW events, meaning only event rates affect our number of events. In reality, all these factors will impact the number of BNS detections. LVK detectors have a duty cycle of 60-75\%\footnote{See \url{https://observing.docs.ligo.org/plan/}.}. Fermi has a field of view that encompasses the whole sky that's not obstructed by the Earth\footnote{See \url{https://fermi.gsfc.nasa.gov/science/instruments/table1-2.html}} with a duty cycle of 60\%\footnote{See \url{https://fermi.gsfc.nasa.gov/ssc/observations/types/grbs/}}. Swift has a smaller field of view, about 20\% the sky, with a duty cycle of 78\% {\cite{Lien:2016zny}}. Our taking one year into account with 100\% duty cycles is justified by the fact that observing runs span multiple years.

\subsection{Einstein Telescope}

As for the LVK case, the $\aMzero, \aTzero, H_0$ posterior for the Einstein Telescope is described by equation \ref{eq:posterior short version}. Since all the simulated ET events have an associated gamma-ray burst, the only likelihood that appears is $\mathcal{L_{\rm GRB}}$, though its form differs to the above. In this case, the GRB detection probability $p(det_{\rm GRB})$ term appearing in eq.~\ref{eq:GRBpost} then depends on flux, which in turn depends on redshift and luminosity, or 
$$
 p(det_{\rm GRB}|H_0) = \iint p(det_{\rm GRB}|F, L, z_{\rm obs}, H_0)p(F|L, z_{\rm obs}, H_0) p(L) \,\dd F \,\dd L \,.
$$
 The $p(F|L, z_{\rm obs})$ term becomes a $\delta$-function, $\delta(F-\hat{F}(L, z_{\rm obs}, H_0))$ which collapses the $F$ integral to $\hat{F}$, given by the expression in eq.~\ref{eq:flux}. 
%
The likelihood in eq.~\ref{eq:full posterior} is further modified to account for the fixed inclination and masses: now GW data only depends on the true GW distance as given by $\aMzero$ and $H_0$. The ET likelihood then is:
\begin{align}
\label{eq:ET likelihood}
    \mathcal{L} \propto \frac{1}{\beta(\aMzero, H_0)} \int_L \int_{\Delta t_{\rm e} = -1s}^{\Delta t_{\rm e} = 10s} &p(x_{\rm GW} | \hat{d}_{\rm GW}(z_{\rm obs}, \aMzero, H_0))p(x_{\Delta t} | \hat{\DTa}(\DTe, z_{\rm obs}, \aTzero, H_0))
    \\ \nonumber &p(det_{\rm GRB}|\hat{F}(L, z_{\rm obs}, H_0))p(\DTe)p(z_{\rm obs}|H_0)p(L) \,\dd\DTe \,\dd L \;.
\end{align}
As in eq.~\ref{eq:full posterior}, we still marginalise over the nuisance parameter $\DTe$. In order to calculate selection effects $\beta(\aMzero, H_0)$, the observational horizon of the Einstein Telescope $d^{\rm thr}_{\rm GW}$ was computed as described above. Additionally, the flux cut imposed on GRBs needs to be accounted for: we only consider events with a joint GW and GRB detection, meaning our mock data will be distributed below the aforementioned cutoff at $d_{\rm L} \sim 3000~$Mpc. This cutoff correspond to a  redshift of $\sim 0.5$, which is below current and future galaxy surveys capabilities \cite{LSST:2008ijt, EUCLID:2011zbd, DESI:2016fyo, SDSS:2023tbz}, allowing us to assume each event has a detected corresponding redshift.
Selection effects are computed according to \cite{Mandel:2018mve}: 
\begin{align}
    \hspace*{-0.4cm}
    \beta(\aM, H_0) = \int_0^{z_{\rm max}} \int_L p(det_{\rm GW}|z, H_0, \aMzero) p(det_{\rm GRB}|\hat{F}(L, z, H_0)) p(z|H_0) p(L)\,\dd L \,\dd z \; ,
\end{align}
where $z_{\rm max}$ is the maximum redshift we consider. Here we use $z_{\rm max} = 2$ since that is much larger than the observational horizon for both GWs and GRBs for the entire prior volume considered. $p(det_{\rm GW}|z, H_0, \aMzero)$ and $p(det_{\rm GRB}|\hat{F}(L, z, H_0))$ are detection probabilities of gravitational waves and gamma-ray bursts, respectively:
\begin{align}
      &p(det_{\rm GW}|z, H_0, \aMzero) = \int_0^{d^{\rm thr}_{\rm GW}} p(x_{\rm GW}|\hat{d}_{\rm GW}(z, \aMzero, H_0)) \, \dd {x_{\rm GW}}
     \\ &p(det_{\rm GRB}|\hat{F}(L, z, H_0)) =  \Theta(\hat{F}(L, z, H_0) - F_{\rm thr}) \; .
\end{align}
The GRB detection probability is a heaviside function $\Theta$: it equals $1$ for all fluxes above the threshold and $0$ for all fluxes below it.
The priors used for the analysis in this scenario are the same as the LVK `GR + Cosmology' scenario given in Table~\ref{tab:priors_H_alphas}, alongside the Gaussian intrinsic luminosity prior mentioned above. 

\section{Results}
\label{sec:Results}

\subsection{LIGO--Virgo--KAGRA Network}
\label{ssec:LVK results}
\begin{figure}[h!]
    \centering
    \includegraphics[scale=0.7]{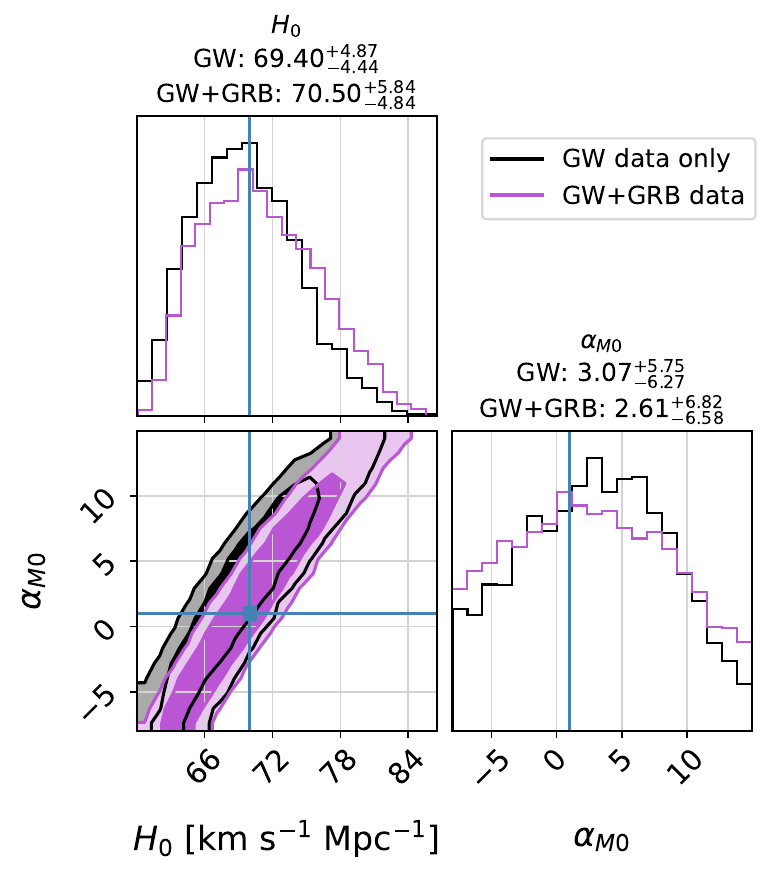}
    \caption{Comparison of $10$ events with and without GRB information. The purple contours are obtained using the restricted inclination range when obtaining the posterior for the one event with GRB. The black contours show the same posterior without GRB information. Levels show $1$ and $2 \sigma$ contours. The blue lines correspond to the injected values of $H_0 = 70\, \text{km s}^{-1}\text{Mpc}^{-1}$ and $\aMzero =1$.}
    \label{fig:GRBcomparison}
\end{figure}

\begin{figure}
    \centering
    \includegraphics[width=0.7\linewidth]{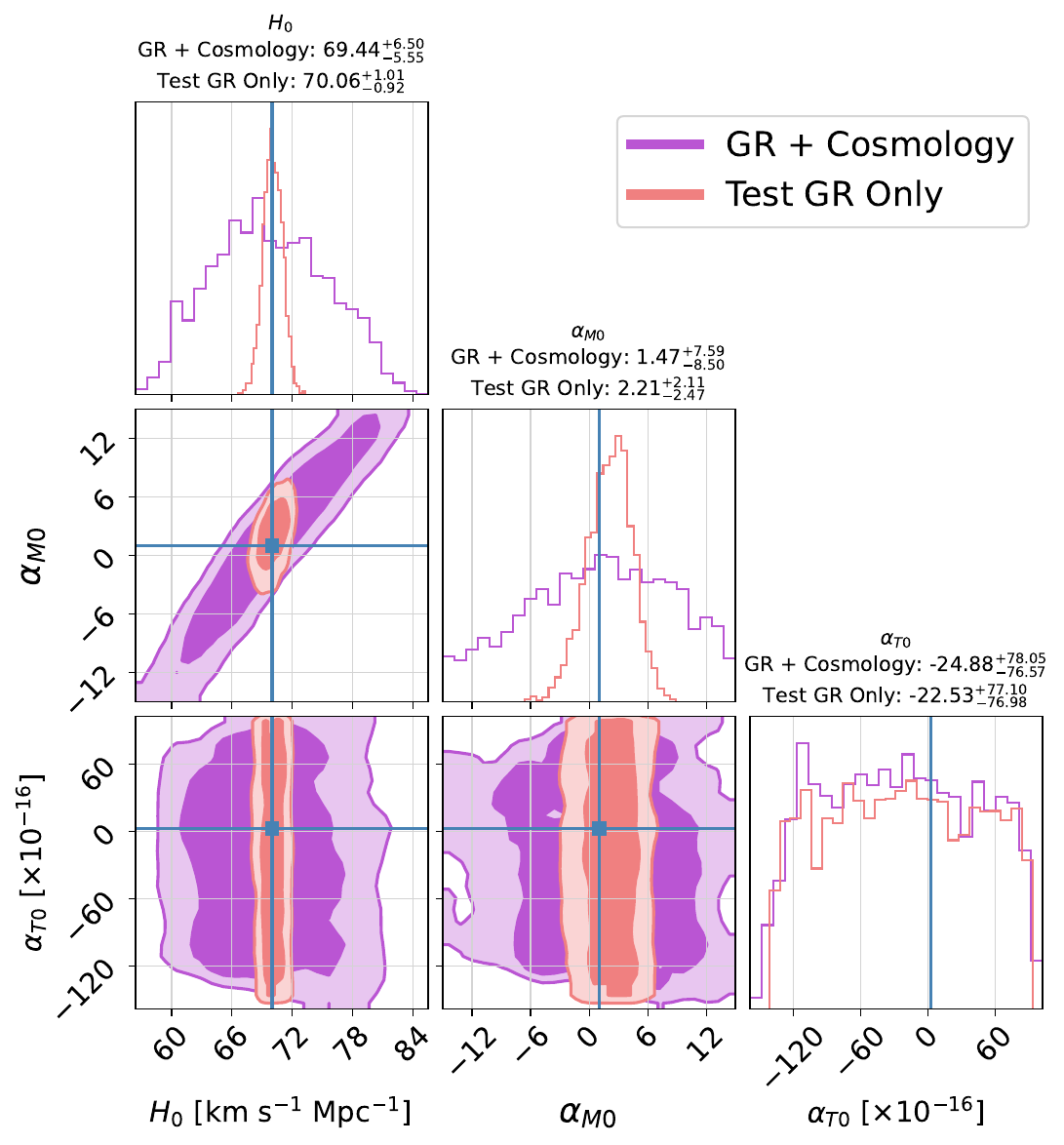}
    \caption{Posteriors for 10 events (one with GRB) with flat priors for all parameters (both testing GR and inferring Cosmology, in purple) and with a narrow Gaussian prior on $H_0$ (testing GR only, in orange). Levels show $1$ and $2 \sigma$ contours. The blue lines correspond to the injected values of $H_0 = 70\, \text{km s}^{-1}\text{Mpc}^{-1}$, $\aMzero =1$, and $\aTzero = 2.699 \times 10^{-16}$.}
    \label{fig:10ev_comparison}
\end{figure}
We first present results for the Hubble parameter using the 10 mock bright sirens in a case where we are in the GR limit. Simulating these events according to GR -- and hence fixing $\aMzero = \aTzero = 0$ -- yields $H_0 = 69.32^{+1.3}_{-1.7}\, \text{km s}^{-1}\text{Mpc}^{-1}$ (68\% confidence interval). This result is illustrative of what can be achieved if modified gravity models are disfavoured by other probes i.e. using $\aMzero = 0$ as a prior. This level of precision is almost on par with the CMB and SH0ES measurements of $H_0 = 67.4 \pm 0.5\, \text{km s}^{-1}\text{Mpc}^{-1}$ \cite{Planck:2018vyg} and $H_0 = 74.03 \pm 1.42\, \text{km s}^{-1}\text{Mpc}^{-1}$ \cite{Riess:2019cxk} (both 68\% confidence interval) and would favour one value over the other leading towards the end of the Hubble tension. In reality, if the universe is not governed by GR and the analysis is carried out this way we will encounter bias in our result: in our fiducial cosmology with $\aMzero = 1$, if we wrongly assume GR we obtain $H_0 = 68.56^{+1.25}_{-1.65}\, \text{km s}^{-1}\text{Mpc}^{-1}$ (1$\sigma$ uncertainties). This specific result is not heavily biased due to the small number of events; however one must proceed with caution as bias accumulates if more incorrectly analysed data points are added.

For our fiducial cosmology the limits obtained on the Hubble constant and $\aMzero$ (by evaluating eq.~\ref{eq:full posterior} for the 10 mock bright sirens) are shown in Figure~\ref{fig:GRBcomparison}. Here we show the effect of using the presence of a GRB for one of the events in order to break the inclination--distance degeneracy (in purple) and GW data only (in black).  $\aTzero$ does not appear in this plot, as the GW--only posterior offers no information on the speed of propagation of GWs due to the lack of GRBs.
An improvement on the Hubble constant is visible compared to the current LVK result of $70^{+12}_{-8}\, \text{km s}^{-1}\text{Mpc}^{-1}$, whose $1\sigma$ uncertainty can be reduced to approximately $\pm 5\, \text{km s}^{-1}\text{Mpc}^{-1}$ with as few as 10 bright siren events. A weakening in errors compared to the GR results stated above is evident, though not surprising as the $H_0$ dependence on the Horndeski parameters widens the posterior. 
Despite this,  combining with dark sirens as well, gravitational waves stand to shed light onto the Hubble tension before next generation detectors begin their runs.

Adding GRB data to a single event out of 10 produces a small shift in the 2D posterior, but ultimately does not tighten the constraints. As mentioned in Section~\ref{ssec:Distances in MG}, effects of $\aM$ are cumulative over redshift, so the wide error bars on $\aMzero$ are not surprising since these events are close-by. An error of $\pm 6$ at 1$\sigma$ is not competitive with the latest result of $\aMzero = 0.98 \pm 0.89$ from DESI \cite{DESI:2024yrg}.

The purple contours, reproduced in Figure~\ref{fig:GRBcomparison} as well, represent the scenario where we wish to constrain both GR and cosmology, where flat priors are used for $H_0$, $\aMzero$, and $\aTzero$\footnote{The bounds on $H_0$ and $\aMzero$ are slightly wider than in Figure~\ref{fig:GRBcomparison} due to marginalisation over $\aTzero$.}.
To improve constraints one either has to look at higher redshift BNSs or introduce more informative priors. The former is not possible with current ground--based detectors, as the horizon for LVK O5 will extend to $240-325$ Mpc at most ($z\sim 0.05-0.07$). We applied more informative priors on $H_0$ in our second scenario (constraining GR only) to exploit its strong correlation with $\aM$, which is evident in Figure~\ref{fig:GRBcomparison}. By adjusting the $H_0$ prior to a narrow Gaussian centred at our injected value and 1$\sigma$ errors of $1\, \text{km s}^{-1}\text{Mpc}^{-1}$, we obtain the orange contours in Figure~\ref{fig:10ev_comparison}. 
The consequent result of $H_0 = 69.44^{+1.01}_{-0.92}\, \text{km s}^{-1}\text{Mpc}^{-1}$ is completely driven by the prior, however the correlation with $\aMzero$ provides a $65-70 \%$ improvement in the error bars of $\aMzero$, constrained to be $\aMzero = 1.47^{+2.11}_{2.47}$. On the other hand, the $\aTzero$ prior doesn't present any improvement due to the very weak correlation with $H_0$ as per eq.~\ref{eq:SoG}; most of the error budget for this quantity is unsurprisingly driven by the emission time delay. Nevertheless, the result for this quantity is one order of magnitude tighter than GW170817, which we used as a prior. Though posteriors improved, both $\aTzero$ and $\aMzero$ are still consistent with GR (recall that our injected fiducial cosmology is non-GR, see Section~\ref{sec:Simulation}): hence it is very likely that using exclusively bright sirens will not bring conclusive results with LVK data.

\subsection{Einstein Telescope}
\label{ssec:ET results}

As with the LVK case, we carry out a one dimensional GR analysis of $H_0$ (i.e. fixing $\aMzero$ and $\aTzero$ to 0 as opposed to marginalising over them) using the 150 ET events with distances with 
Gaussian uncertainties. We find $70.32^{+ 0.41}_{-0.40}\, \text{km s}^{-1}\text{Mpc}^{-1}$ which would be the most precise result on this quantity for any probe to date\footnote{An analysis assuming GR in our modified gravity universe yields $62.15^{+ 0.37}_{- 0.35}\, \text{km s}^{-1}\text{Mpc}^{-1}$, which is a highly biased result -- one needs to be very careful not to disregard modified gravity theories.} (though naturally by then other probes will also have advanced).

\begin{figure}[h]
    \centering
    \includegraphics[scale=0.6]{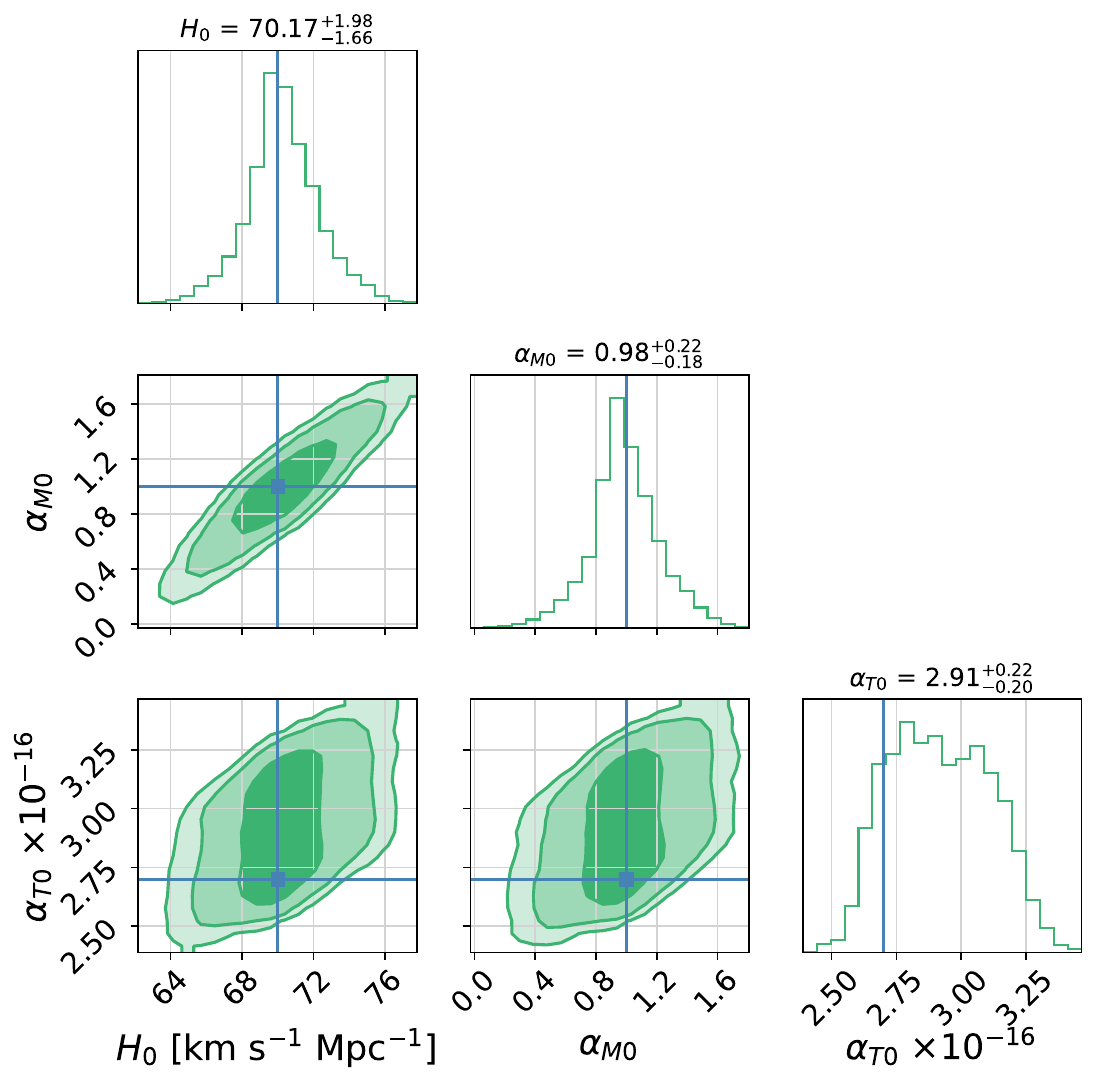}
    \caption{Posteriors for $150$ ET events with GRBs, using flat priors for all inferred quantities. The contours represent $1$, $2$ and $3 \sigma$ levels and results above each panel have errors corresponding to a 68.3 \% credible interval. The blue lines correspond to the injected values of $H_0 = 70\, \text{km s}^{-1}\text{Mpc}^{-1}$, $\aMzero =1$, and $\aTzero = 2.699 \times 10^{-16}$.}
    \label{fig:ETresult}
\end{figure}

Using the likelihood in equation~\ref{eq:ET likelihood} and the ET events, we obtain the 3D posterior shown in Figure~\ref{fig:ETresult}. The priors in this case are flat for all three parameters and the contours indicate 1, 2, and $3 \sigma$ levels. A reduced inclination range is used here as all events have an associated GRB -- we will see the impact this has on the result later. The strong correlation between $H_0$ and $\aMzero$ is once again evident in this plot, this is a \textit{positive} correlation, as expected from eq.~\ref{eq:dGW}: increasing (decreasing) $H_0$ requires increasing (decreasing) $\aMzero$ to obtain the same value of $d_{\rm GW}$. Using many events we see that results start being sensitive to the slight correlation between $H_0$ and $\aTzero$ as well, once again positively tilted as expected from eq.~\ref{eq:SoG}.

The Hubble parameter result is $H_0 = 70.17^{+1.98}_{-1.66} \, \text{km s}^{-1}\text{Mpc}^{-1}$ for a 68.3\% credible interval. This is only marginally less precise than the Planck value
and comparable to the Cepheids result. As in the LVK case, bounds are once again wider than the GR limit due to the additional modified gravity parameters.

Showing a large improvement compared to LVK constraints, a result of this type could disagree with one of the two values for $H_0$ that are in tension. The disagreement could be around $2 \sigma$, which on its own would not exclude either value, however added to current and future GW cosmological analyses may solve the tension. 

For the chosen fiducial model, both $\aTzero$ and $\aMzero$ in Figure~\ref{fig:ETresult} are incompatible with GR at $3 \sigma$. This result means that deviations from GR, if present, could be detected with next generation detectors within as soon as one year of observations.

\begin{figure}[h!]
    \centering
    \includegraphics[scale=0.7]{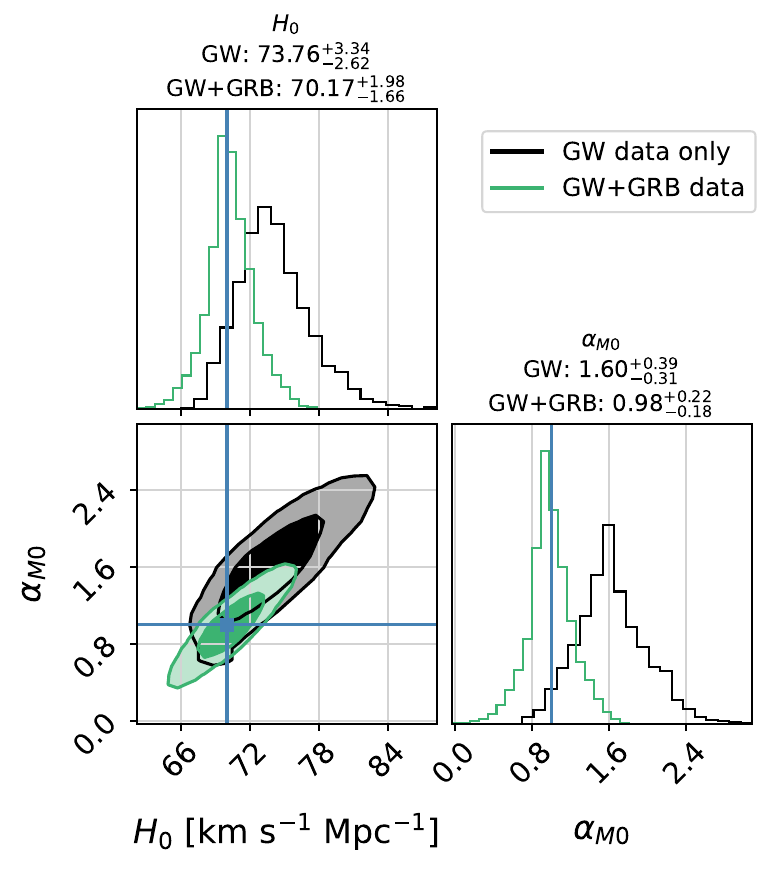}
    \caption{Posterior for 150 ET events, comparing cases with and without GRB information to restrict the inclination range. The green contours indicate the former (giving a distance fractional error of $7\%$), while the black contours represent the latter (error of $20\%$). The blue lines correspond to the injected values of $H_0 = 70\, \text{km s}^{-1}\text{Mpc}^{-1}$ and $\aMzero =1$.}
    \label{fig:ET_comparison}
\end{figure}

Figure~\ref{fig:ET_comparison} is analogous to Figure~\ref{fig:GRBcomparison}, showing the impact of reducing the inclination range (remember this is possible as all 150 events have corresponding GRBs) on the $H_0$--$\aMzero$ posterior. There is a visible difference between using the GRB data to marginalise over smaller inclination range (green contour, same as in Figure~\ref{fig:ETresult}) compared to marginalising over the whole range (black contour). In fact, not including this information results in error bars that are twice as large for both $H_0$ and $\aMzero$. This highlights the importance of focusing on EM follow-up in the future in order to pin down inclination from GRBs and afterglows.

\begin{figure}[h!]
    \centering
    \includegraphics[scale=0.6]{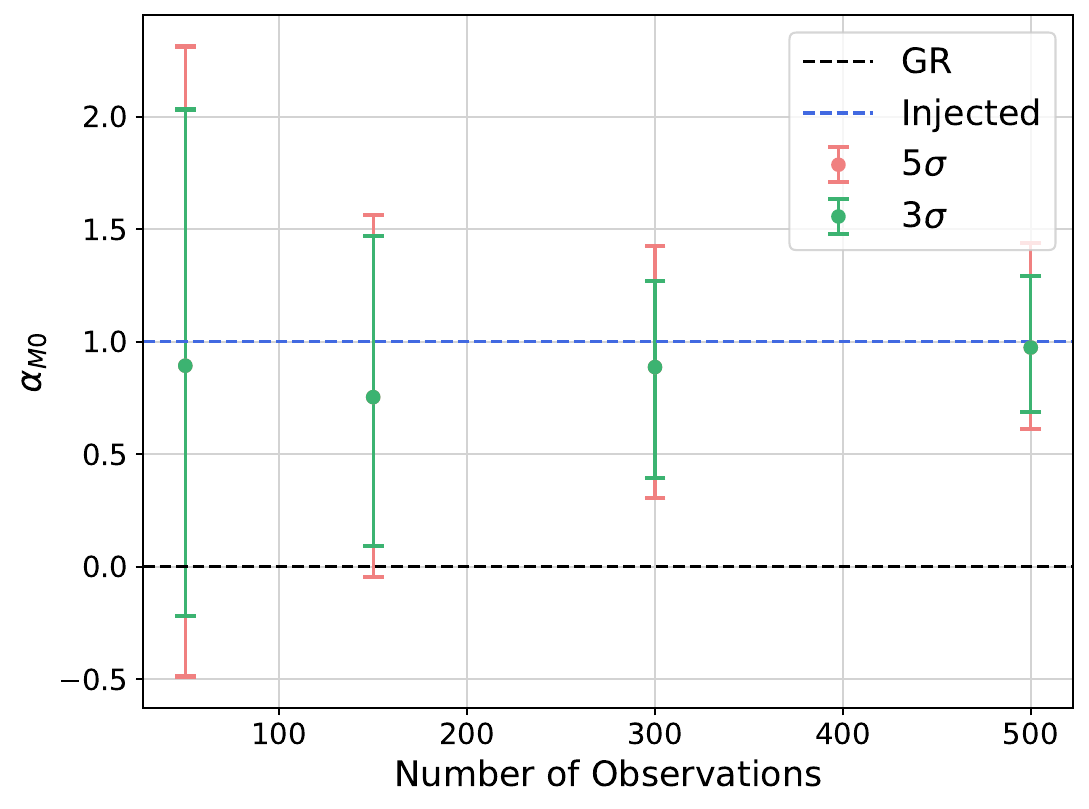}
    \caption{$\aMzero$ error bars for 50, 150, 300, and 500 ET events. Both $3$ and $5\sigma$ uncertainties are shown, in green and orange respectively. The dashed horizontal lines correspond to GR (black) and the injected value $\aMzero=1$ (blue).}
    \label{fig:different_nevents}
\end{figure}

In order to investigate the full capabilities of ET we performed additional three-dimensional runs (i.e. inferring $\aMzero$, $\aTzero$, and $H_0$ simultaneously) with 50, 300, and 500 events (all with GRB and restricted inclination prior). These were carried out according to the method described for the ET case.  In this context, increasing the number of events can be seen as analogous to selecting a kilonova intrinsic luminosity from the brighter end of the distribution in {\cite{Ghirlanda:2016ijf, Wanderman:2014eza}}. We specifically focus on the marginal result for $\aMzero$, showing results of these analyses in Figure~{\ref{fig:different_nevents}}. As expected the error bars shrink according to $\sim \sqrt{N}$. We see that for 300 events already there is a $5\sigma$ deviation from GR, an exciting prospect considering the expected rate of detections of ET. 

\section{Conclusions}

In this paper we have presented a prediction of the role bright sirens (BNSs) will have in cosmology and tests of GR over the planned next LVK runs and beyond. We have produced forecasts for the Hubble constant and the functions $\aM(z)$ and $\aT(z)$, which parametrise changes in the effective gravitational coupling strength 
and the speed of propagation of gravitational waves in Horndeski gravity theories. We showed the full Bayesian framework for both LVK O5 and ET scenarios, assuming only a fraction of LVK events have an associated GRB. For the LVK scenario, we have shown that using ten bright sirens alone gives $H_0 = 69.44^{+6.50}_{-5.55} \, \text{km s}^{-1}\text{Mpc}^{-1}$ for an injected value of $70\, \text{km s}^{-1}\text{Mpc}^{-1}$, which is a reduction in errors on current GW-inferred $H_0$ by a half.
The $\aMzero$ posterior, however, has a large degree of uncertainty due to the proximity of the BNSs. If the $H_0$ tension is resolved, the measurement of $\aMzero$ improves greatly, yielding $\aMzero = 1.47^{+2.11}_{-2.47}$. Though the value is comparable with current dark sirens constraints of $\aMzero = 1.5^{+2.2}_{-2.1}$ \cite{Chen:2023wpj} (and further constraints found in \cite{Mancarella:2022cgn, Leyde:2022orh} ), it 
remains mildly less constraining than the DESI bound
of $\aMzero = 0.98 \pm 0.89$ (1$\sigma$ level) \cite{DESI:2024yrg}. Despite these constraints not being comparable, EM probes and GWs need to be studied separately as well as jointly, as possible discrepancies between the two could be an indication that the wrong theory is being used.

A word of caution is necessary: the results here were obtained under a particular ansatz for the redshift dependence of $\aM(z)$ and $\aT(z)$. Modifying the ansatz could change our results, though the change is expected to be somewhat marginal (see e.g. \cite{Baker:2020apq, Seraille:2024beb}) especially in the LVK range where modified gravity effects are weak. As previously mentioned, we selected one widely-used parametrisation to facilitate comparison with the literature. 

In Section~\ref{ssec:ET results} we showed that third generation GW detectors could detect deviations from GR at 3$\sigma$ confidence (with accuracy scaling with the number of events) through $\aMzero$ and $\aTzero$ within the first year of observations, for our simulated Horndeski universe. However, to do this it is important to have GRB information available to narrow the inclination range when computing the distance posteriors.

Our results indicate that bright sirens alone will not yield further groundbreaking results on GW propagation tests of GR in O4/O5. This may come as a surprise, given that GW170817 was a very powerful event that ruled out some theories of modified gravity. Our findings hence highlight the necessity to develop alternatives such as the dark sirens method, though this comes with a caveat: the process of assigning a redshift distribution to a non-counterpart event often provides weak posteriors. This problem stems from the lack of precise localisation data and the incompleteness of the galaxy catalogue -- though efforts are being made to overcome this \cite{Mukherjee:2020hyn, Finke:2021aom, Dalang:2023ehp, Leyde:2024tov} -- and uncertainties in source frame mass distribution which can greatly impact results \cite{Mastrogiovanni:2021wsd}. The ever-growing number of detections will mitigate this issue, though a single bright siren event will remain more constraining than a single dark siren event. Of course, both methods should be pursued in parallel, and optimal constraints will come from their joint analysis. 

Though tests on cosmological scales will not be highly impacted by bright sirens, these only probe GW \textit{propagation}. Waveform consistency checks carried out by the LIGO--Virgo--KAGRA collaboration mostly probe GW \textit{generation}, which was not discussed in this paper. Though these tests have found no inconsistencies with GR to date \cite{LIGOScientific:2018dkp, LIGOScientific:2019fpa, LIGOScientific:2020tif, LIGOScientific:2021sio}, we wish to highlight that our statement on the role of bright sirens in the future does not include these tests.

Throughout this paper we also have stressed the importance of detecting GRBs in case a bright counterpart is present. Continuing investment is needed in the EM follow-up process and GRB detection both to ensure measurements of redshifts and time delays and to break the inclination--distance degeneracy. A few dedicated projects are already underway to facilitate this task. The Gravitational-wave Optical Transient Observer (GOTO) \cite{Dyer:2024bwm} is specifically built to perform follow-up searches of GRBs from GWs and has automatised pipelines for speed. The recently launched Einstein Probe \cite{Yuan:2022fpj} can observe EM transients in the X-ray band to produce better localisation of a source (though this is not a detection of the initial burst it can still be used to infer inclination). Additionally, a network of GRB sensors on second-generation Galileo satellites was proposed \cite{Greiner:2022hxe}. The advantage of this network would be increased localisation accuracy via triangulation. All of these initiatives are essential and must be continually updated to ensure coordination with third generation GW detectors. Between the end of O5 and ET coming online there are plans to upgrade sensitivity of current detectors, with the A\# era \cite{Fritschel2022}, and to have LIGO Aundha in India begin observations in the 2030s \cite{Saleem:2021iwi, Unnikrishnan:2023uou}. These coinciding advancements would improve the number and accuracy of detections allowing for an accelerated timeline to obtain cosmological results. 

In this work we have focused on tests of gravity with gravitational waves, and emphasised the value of such approaches. Cosmology in the 2030s will bring GW sources into the fold of multi-probe analysis, as is done currently, for example, with baryonic acoustic oscillations, supernovae and CMB \cite{DESI:2024mwx}. Though challenging, this will enable us to leverage the wildly different nature of GW sources and galaxies to obtain new insights on gravity and cosmology.

\acknowledgments

We would like to thank Gareth Cabourn Davies, Charlie Hoy, and Michael Williams for their assistance with \texttt{pyCBC} and \texttt{bilby}, as well as Antonio Enea Romano, Michele Mancarella, and Danièle Steer for useful discussions and comments. The authors are grateful for computational resources provided by the LIGO Laboratory and supported by National Science Foundation Grants PHY-0757058 and PHY-0823459. We additionally thank Simone Mastrogiovanni for reviewing this work. Analyses and plotting were performed using the \texttt{emcee} \cite{Foreman-Mackey:2012any} and \texttt{corner.py} \cite{corner} packages. E.C., T.B. and K.L. are supported by ERC Starting Grant SHADE (grant no.\,StG 949572). T. B. is further supported by a Royal Society University Research Fellowship (grant no.\,URF$\backslash$R$\backslash$231006).

\appendix

\section{Mass Scales in Horndeski Gravity}
\label{app: aT and aM}

Equation~\ref{eq:M_*aT} explicitly shows the relation between $M_*$ and the tensor speed excess parameter $\aT$. When applying an effective field theory approach to the Horndeski action, one can find an explicit dependence of the $d_{\rm GW}/d_{\rm EM}$ ratio on the propagation speed. Here we will show that the consequent impact $\aT$ has on $\aM$ is negligible in our case, and hence the two functions can be inferred independently of each other. 

In the effective field theory of dark energy formalism, one can write $M_* = M_{\rm eff}/c_{\rm T}$, where $M_{\rm eff}$ is a second mass scale that acts as the effective Planck mass (in place of $M_*$). Considering the definition of $\aM$ in eq.~\ref{eq:alphaM definition}, this yields \cite{Romano:2022jeh}:

\begin{align}
    \aM = \frac{1}{H M_*} \frac{d}{dt} M_* = \frac{2}{H} \bigg[\frac{\dot{M}_{\rm eff}}{M_{\rm eff}}-\frac{\dot{c}_{\rm T}}{c_{\rm T}} \bigg] \; .
\end{align}
We can approximate the second term in the bracket as
\begin{align}
    \frac{\dot{c}_{\rm T}}{c_{\rm T}} = \frac{c\dot{\alpha}_{\text{T}}/2}{c(1+\aT/2)} \approx \frac{\dot{\alpha}_{\text{T}}}{2} \; ,
\end{align}
where we have used that current data restrict $\aT\ll 1$. Applying our parametrisation $\aT = \aTzero \Omega_\Lambda(z)/\Omega_{\Lambda 0}$, and using that $\Omega_\Lambda = \Omega_{\Lambda 0} H_0^2/H^2$ for a $\Lambda CDM$ expansion history (which we assume in this work), one obtains 
\begin{align}
    \frac{\dot{\alpha}_{\text{T}}}{2} = \frac{1}{2}\frac{\dot{\Omega}_\Lambda}{\Omega_{\Lambda 0}} =  H_0^2  \frac{1}{2} \frac{d}{dt}\bigg[ \frac{1}{H^2}\bigg] = H_0^2 \bigg[ -\frac{2 \dot{H}}{H^3}\bigg] = - \bigg(\frac{H_0}{H} \bigg)^2 \frac{\dot{H}}{H} \; .
\end{align}
This result is of order $10^{-18}$, and since we are constraining $\aMzero$ up to order $10^{-3}$ this effect can be ignored in our analysis.

\section{Arrival Time Delay Derivation}
\label{app:SoG Derivation}

The following is the complete derivation leading to eq.~\ref{eq:SoG}, which follows the derivation from \cite{Tasson:2020}.

Firstly we define a few quantities which will be used in the computation: the speed difference $\Delta v = v_{\rm GW} - v_{\rm EM}$, emission time delay $\DTe = t_{\rm e, EM} - t_{\rm e, GW}$ and arrival time delay $\DTa = t_{\rm a, EM} - t_{\rm a, GW}$. As for the speed difference, we Taylor expand from equation~\ref{eq:alphaT definition}: $v_{\rm GW} \approx v_{\rm EM}(1+\aT/2) \implies v_{\rm GW}/v_{\rm EM} - 1 = \aT/2$. We change our notation here to be consistent with \cite{Tasson:2020}, however $c_{\rm T} \equiv v_{\rm GW}$ and $c \equiv v_{\rm EM}$.

Taking the comoving path for a photon and a gravitational wave emitted by a binary and equating the two gives
\begin{align}
    \int_{z_{\rm a, EM}}^{z_{\rm e, EM}} \frac{v_{\rm EM}}{H(z')} \,\dd z' = \int_{z_{\rm a, GW}}^{z_{\rm e, GW}} \frac{v_{\rm EM}}{H(z')}\bigg(1+\frac{\Delta v}{v_{\rm EM}}\bigg) \,\dd z' \; ,
\end{align}
where the left hand side is the photon path and the right is the GW's.
Following \cite{Tasson:2020} we redefine the limits of the integrals according to: $\Delta z_{\rm a} = z_{\rm a, GW} - z_{\rm a, EM}$, $\Delta z_{\rm e} = z_{\rm e, GW} - z_{\rm e, EM}$, $z = z_{\rm e, GW}$ and $z_{\rm a, EM} = 0$:
\begin{align}
    \int_0^{z-\Delta z_{\rm e}}  \frac{1}{H(z')} \,\dd z' &= \int_{\Delta z_{\rm a}}^z \frac{1}{H(z')} \bigg( 1 + \frac{\Delta v}{v_{\rm EM}} \bigg) \,\dd z'
\end{align}
This allows us to split the above integrals:
\begin{align}
    \nonumber \int_0^z \frac{1}{H(z')} \,\dd z' + \int_z^{z-\Delta z_{\rm e}} \frac{1}{H(z')} \,\dd z' &= \int_0^z \frac{1}{H(z')} \bigg( 1 + \frac{\Delta v}{v_{\rm EM}} \bigg) \,\dd z' - \bigg( 1 + \frac{\Delta v}{v_{\rm EM}} \bigg) \int_0^{\Delta z_{\rm a}} \frac{1}{H(z')} \,\dd z'
    \\ \nonumber \int_z^{z-\Delta z_{\rm e}} \frac{1}{H(z')} \,\dd z' &= \int_0^z \frac{1}{H(z')} \frac{\aT(z')}{2} \,\dd z' - \bigg( 1 + \frac{\Delta v}{v_{\rm EM}} \bigg) \int_0^{\Delta z_{\rm a}} \frac{1}{H(z')} \,\dd z'
    \\ \int_z^{z-\Delta z_{\rm e}} \frac{1}{H(z')} \,\dd z' + \int_0^{\Delta z_{\rm a}} \frac{1}{H(z')} \,\dd z' &= \frac{\aTzero}{2 H_0} \int_0^z \frac{1}{E(z')}\frac{\Omega_\Lambda(z')}{\Omega_{\Lambda0}} \,\dd z' - \frac{\Delta v}{v_{\rm EM}} \int_0^{\Delta z_{\rm a}} \frac{1}{H(z')} \,\dd z' \; .
\end{align}
Here we used $v_{\rm GW}/v_{\rm EM} - 1 = \aT/2$, as well as cancelled the integrals from 0 to $z$ on both sides. The second term on the right is negligible as it is a minuscule distance which is much smaller than the first term on the right. The integrals on the left also span a small range, meaning they can be approximated to redshift intervals:
\begin{align}
    \int_z^{z-\Delta z_{\rm e}} \frac{1}{H(z')} \,\dd z' + \int_0^{\Delta z_{\rm a}} \frac{1}{H(z')} \,\dd z' \approx \frac{\Delta z_{\rm a}}{H_0} - \frac{\Delta z_{\rm e}}{H(z)} \approx \DTa - (1+z) \DTe \; ,
\end{align}
where redshift intervals are converted to time intervals following $dt = -\frac{dz}{H(z)(1+z)}$. After performing these approximations, the above expression becomes:
\begin{align}
    \DTa - (1+z) \DTe = \frac{\aTzero}{2 H_0} \bigg[\int_0^z \frac{1}{E(z')}\frac{\Omega_\Lambda(z')}{\Omega_{\Lambda0}} \,\dd z'\bigg] \; ,
\end{align}
which can be easily rearranged to obtain~\ref{eq:SoG}.

\section{\texttt{bilby} Specifics}
\label{app:bilby}

In order to obtain our simulated data for the LVK scenario we used the \texttt{IMRPhenoPv2\_NRTidal} waveform combined with the reduced order quadrature (ROQ) method \cite{roq_paper_1, roq_paper_2} to limit computational costs. Phase, time of coalescence, tidal deformability parameters, spins, and sky location were fixed. The priors used for the non-fixed parameters are shown in Table~\ref{tab:bilby_priors}. These are chirp mass $\mathcal{M} = (m_1 m_2)^{3/5}/(m_1 + m_2)^{1/5}$ ($m_1, m_2$ are the component masses of the binary), mass ratio $q = m_2/m_1$, GW distance $d_{\rm GW}$, and inclination $\iota$. All posteriors were obtained for a no-noise scenario with LIGO Hanford, LIGO Livingston and Virgo detectors being online. 

\begin{table}{H}

\begin{center}
\begin{tabular}{ !{\vrule width 0.5mm} c | c | c !{\vrule width 0.5mm} } 
\hlineB{3}
Parameter & \texttt{bilby} prior & Range \\
\hline
$\mathcal{M}$ & UniformInComponentsChirpMass & $[0.92, 1.7] M_\odot$ \\ 
$q$ & UniformInComponentsMassRatio & $[0.25, 1]$\\
$d_{\rm GW}$ & UniformComovingVolume & $[10, 1000]$ Mpc \\
$\iota$ & Sin & $[0, \pi]$ \\ 
\hlineB{3}
\end{tabular}
\end{center}
\caption{Priors used to obtain the GW distance posterior used for the O4 and O5 analysis.}
\label{tab:bilby_priors}
\end{table}

\bibliographystyle{JHEP}
\bibliography{main}

\end{document}